\documentclass[%
 reprint,
 amsmath,amssymb,
 prb,
]{revtex4-2}

\usepackage{graphicx}
\usepackage[rgb]{xcolor}
\definecolor{dgreen}{rgb}{0.0,0.7,0.0}

\usepackage[unicode, colorlinks=true, allcolors=blue]{hyperref}
\numberwithin{equation}{section}
\usepackage{braket}
\newcommand\rotpm{\rotatebox[origin=c]{180}{$\pm$}}
\newcommand\eqDef{\mathrel{\overset{\makebox[0pt]{\mbox{\tiny def}}}{=}}}
\DeclareUnicodeCharacter{2212}{-}

\usepackage{dcolumn}
\usepackage{bm}

\begin{document}

\preprint{APS/123-QED}

\title{Exciton-exciton and exciton-charge carrier interaction \\and the exciton collisional broadening \\in GaAs/AlGaAs quantum wells}

\author{B.F. Gribakin}
 \email{bgribakin@gmail.com}
\author{E.S. Khramtsov}%
\author{A.V. Trifonov}
 \altaffiliation[Also at ]{Experimentelle Physik 2, Technische Universit\"at Dortmund, Otto-Hahn Stra{\ss}e 4a, Dortmund, 44227, Germany}
\author{I.V. Ignatiev}
\affiliation{%
 Spin Optics Laboratory, St. Petersburg State University\\
 Ulyanovskaya 1, 198504, St. Petersburg, Russia
}%

\date{\today}

\begin{abstract}
Wave functions of heavy-hole excitons in GaAs/Al$_{0.3}$Ga$_{0.7}$As square quantum wells (QWs) of various widths are calculated by the direct numerical solution of a three-dimensional Schr\"odinger equation using a finite-difference scheme. These wave functions are then used to determine the exciton-exciton, exciton-electron and exciton-hole fermion exchange constants in a wide range of QW widths ($5 - 150$~nm). Additionally, the spin-dependent matrix elements of elastic exciton-exciton, exciton-electron and exciton-hole scattering are calculated. From these matrix elements, the collisional broadening of the exciton resonance is obtained within the Born approximation as a function of the areal density of excitons, electrons and holes respectively for QW widths of 5, 15, 30 and 50~nm. The obtained numerical results are compared with other theoretical works. 
\end{abstract}

\maketitle




\section{Introduction} 
\label{sec:introduction}

At low temperatures, the optical spectra of high-quality semiconductor nanostructures are dominated by excitonic transitions, which are observed as homogeneously broadened resonances. The energies of these resonances are generally well predicted by  modern theoretical approaches~\cite{ivchenko2005book, khramtsov2016}. However, another important characteristic of such resonances is the broadening, which consists of the radiative and non-radiative contributions. While the radiative broadening of a given excitonic resonance may be calculated with reasonable accuracy if the wave function is known~\cite{andreani1991-radiative-lifetime, d'andrea1998-radiative-lifetime, khramtsov2016, grigoryev2016, khramtsov2019}, modelling the non-radiative broadening is a more difficult problem to tackle, because many possible interactions may contribute to it. 

The non-radiative broadening of an exciton resonance is determined by the interactions between the exciton and other quasiparticles present in  a high-quality heterostructure, such as other excitons, charge carriers or phonons. The study of the non-radiative broadening is one of the possible routes to understanding these interactions. In general, inter-particle interactions give rise to optical nonlinearities. These nonlinearities allow the photons to interact, making the concept of optical computations possible. The various exciton-quasiparticle interactions are widely studied in heterostructures with quantum wells (QWs)~\cite{butov2002, voros2009, butov2015, butov2021}, microcavities~\cite{kavokin2017microcavities}, two-dimensional structures~\cite{shelykh2017, erkensten2021, efimkin2021}, and many other systems~\cite{other1, other2, other3}. 

A common and well studied interaction is the exciton-phonon interaction~\cite{spectorLeeMelman1986e-ph, grigorchuk1999e-ph, gopal2000e-ph, zhao2002e-ph, thranhardt2003e-ph, zhao2004e-ph, poltavtsev2014e-ph}. In this case, the concentration of phonons is determined by the sample's temperature. However, at low temperatures the most important types of quasiparticles interacting with optically active excitons are the free carriers and  other excitons, especially non-radiative excitons with large in-plane wave vectors exceeding the wave vector of light. Although experimental studies of these interactions have been conducted for already three decades~\cite{honold1989, deveaud1991, kaindl2003, szczytko2004, deveaud2005PRB, kaindl2009, trifonov2015, beck2016, kurdyubov2021arXiv}, important characteristics of these interactions, such as the physical mechanisms and the scattering cross-section, are still not uniquely determined. The problem lies in the densities of the quasiparticles, which cannot be determined directly from experiments with sufficient accuracy. Indeed, when measuring concentrations through absorption one is faced with uncertainties in the absorption coefficient and more importantly in the efficiency of quasiparticle generation. When using luminescence to determine the exciton concentrations, uncertainties in the efficiency of photon emission get in the way~\cite{voros2009}. 

Then there is also the problem of non-radiative excitons with large in-plane wave vectors~\cite{trifonov2015, assmann2020, kurdyubov2021arXiv}. In heterostructures, the density of the non-radiative excitons strongly depends not only on the experimental conditions of optical excitation, but also on the quality of the heterostructure. In high-quality structures with quantum wells, the lifetime of the non-radiative excitons reaches tens of nanoseconds~\cite{trifonov2015}, therefore their areal density can exceed that of radiative excitons by orders of magnitude. Apart from these excitons, free carriers may also be created. As a result, a reservoir with a mixture with non-radiative excitons and free carries is formed. The rich dynamics of quasiparticles in this reservoir with exciton formation and dissociation further complicate the problem~\cite{deveaud2005PRB, deveaud2005chemPhys, kurdyubov2021arXiv}.

It should be noted that interactions between quasiparticles produce not only homogeneous broadening, but also lineshifts. However, experiments show that the magnitude of the lineshift corresponding to, e.g, the exciton-exciton interaction, is smaller than the broadening, and is not as sensitive to the concentration~\cite{voros2009}. Therefore we consider the collisional broadening a more convenient quantity in experiments, and do not discuss the lineshifts in this work.

There exists a large amount of theoretical works dedicated specifically to exciton-exciton and exciton-carrier interactions in semiconductor nanostructures. 
In the early works, the spin degrees of freedom were neglected entirely~\cite{haug1976, feng1987}. Later works treated excitons as elementary bosons~\cite{hiroshima1989}, which also proved to be an incomplete treatment. Some of the first works to incorporate the composite character of excitons were conducted by V.~May, F.~Boldt and K.~Henneberger~\cite{dense_gas_1, dense_gas_2}. The importance of interexciton exchange interactions was first suggested in the work~\cite{amand1997}. The collisional broadening of 2D heavy-hole excitons due to exciton-exciton interactions was calculated by C.~Ciuti {\it et. al}~\cite{ciuti1998}. The broadening of neutral and charged excitons due to collisions with electrons was calculated by the group of E.~Cohen~\cite{cohen_neutral&charged_exciton-electron2003}, although a somewhat different approach was used. Both of these works accounted for the spin degrees of freedom and also reported on the dominant role of exchange interactions. Exciton-carrier scattering was also studied in~\cite{ouerdane2008}, however their study was focused on carrier-assisted radiative recombination. Some attempts were also made to introduce an effective potential which would account for the exchange interaction~\cite{schindler2008}. Through these works it became clear that, when discussing exciton interactions, excitons must not be treated as elementary bosons. 

However, as first pointed out by S.~Okumura and T.~Ogawa~\cite{okumura_ogawa2001bosonization}, the previous attempts at this type of treatment were incomplete. They have discovered that in approaches such as in~\cite{ciuti1998} (i.e. the Hartree-Fock approximation), although the fermion composition of the exciton is taken into account, some terms are missed, which leads to an underestimate of the interaction strength, and hence of the broadening as well. M.~Combescot and colleagues have stressed the importance of a proper treatment of the exciton-exciton interaction, and have built a new ``coboson'' theory to describe it (see, for example~\cite{combescot2008}). Their work was expanded upon in a paper by M.~Glazov {\it et. al}~\cite{glazov2009}, in which an attempt has been made to simplify the approach of~\cite{combescot2008}. 

In this work, we present a theoretical study of exciton-exciton and exciton-carrier scattering in GaAs/Al$_{0.3}$Ga$_{0.7}$As square QWs and their effect on the non-radiative broadening of the heavy-hole exciton resonance. The QWs are assumed to be ideal and finite in depth. Microscopic calculations of the exciton envelope wave functions are carried out. Using these wave functions, we determine the exciton-exciton, exciton-electron and exciton-hole exchange constants with respect to QW width. The wave functions are used to calculate the broadening due excitons colliding with other excitons, as well as charge carriers. We exploit the relatively simple model introduced in the paper by C.~Ciuti \textit{et. al}~\cite{ciuti1998}, which does not account for exciton correlation effects and misses some interaction terms~\cite{okumura_ogawa2001bosonization}, but nevertheless gives reliable estimates for exciton-exciton collisional broadening and correctly reproduces the spin scattering channels. In it's original form, the model of Ref.~\cite{ciuti1998} is limited to the treatment of strictly 2D excitons. We generalise it for 3D excitons in ideal square GaAs/Al$_{0.3}$Ga$_{0.7}$As finite QWs, the wave functions of which are calculated as in Refs.~\cite{khramtsov2016,khramtsov2019}. The usage of the numerical wave functions allows us to study the behaviour of the broadening with respect to the width of the QW.  We calculate the exciton line broadening for several QW widths for two limiting cases of spin polarisation, and analyse the QW width dependency of the broadening using two-parameter functions. We also modify the model for the case of exciton-carrier scattering. The lack of any fitting parameters in our model allows the results to be directly compared to experiments, and also provides a way of estimating exciton and carrier concentrations from the broadening.

The paper is organised as follows. In \hyperref[sec:theoretical]{Section 2} we describe the general theoretical framework used in our study of exciton-electron, exciton-hole and exciton-exciton interactions. In \hyperref[sec:matrix_elements_calculation]{Section 3} we elaborate on the numerical methods we used. \hyperref[sec:broadening]{Section 4} is devoted to the problem of density-dependent collisional broadening of the exciton resonance in the three cases discussed. Finally, in \hyperref[sec:conclusions]{Section 5} we draw conclusions and compare our results to literature.

\section{Theoretical model of exciton-carrier and exciton-exciton scattering}
\label{sec:theoretical}
\subsection{Wave functions of excitons and carriers}

We assume an ideal square GaAs/Al$_{0.3}$Ga$_{0.7}$As QW with no defects. We restrict ourselves to the $1s$-like ground state of the heavy-hole exciton, the wave functions of which are numerically obtained according to Ref.~\cite{khramtsov2016}. The numerical envelope wave functions obtained in this way generally model experiments very well when the effective mass approximation is valid~\cite{khramtsov2016, khramtsov2019}. An exciton Hamiltonian is constructed, with the hole in the valence band being described by a Luttinger Hamiltonian. In these calculations, the heavy-hole/light-hole coupling is not taken into account, however for QWs with width $L < 50$~nm the effects of hole subband mixing on exciton wave functions may be neglected for our purposes \cite{grigoryev2021unpublished}. Our estimates reveal that the heavy-hole/light-hole splitting is close to 0.7~meV in the $L = 50$~nm case and quickly grows with decreasing $L$. In wider QWs, our results must be treated as
estimates.

In the absence of heavy-hole/light-hole coupling, the Luttinger Hamiltonian is diagonal. The corresponding exciton Hamiltonian can be written as~\cite{khramtsov2016}
\begin{align} \label{eqn:X_hamiltonian} \nonumber
    \hat{H}_{X} = \frac{\hbar^2 \hat{k}^2_e}{2m_e} &+ \frac{\hbar^2}{2m_{h}^{\perp}} \left( \hat{k}^2_{h,x} + \hat{k}^2_{h,y} \right) + \frac{\hbar^2 \hat{k}^2_{h,z}}{2m_{h}^{_\parallel}}\\
    &+ V_{eh}({\bf r}_e, {\bf r}_h) + V_e(z_e) + V_h(z_h),
\end{align}
where $\hat{k}_e$ is the electron wave vector operator, $\hat{k}_{h,x(y,z)}$ are the hole wave vector operators, ${\bf r}_{e(h)}$ is the electron (hole) coordinate vector, $m_e$ is the electron effective mass, $m_{h}^{_\parallel}$ is the hole mass in the growth direction, $m_{h}^{\perp}$ is the in-plane hole mass. Finally, $V_{eh}$ is the electron-hole Coulomb attraction potential, and $V_{e(h)}$ is the electron (hole) rectangular QW potential. 

Due to the axial symmetry, the Schr\"odinger equation for the 6-dimensional Hamiltonian (\ref{eqn:X_hamiltonian}) is reduced to 3 dimensions. Then the equation is discretised with a finite-difference scheme. Finally, the eigenvalue problem is solved using the Krylov-Schur algorithm~\cite{kressner2005KrylovSchur}, yielding the numerical 1s-like wave function $\phi(\rho_{eh}, z_e, z_h)$, which depends on the in-plane distance between the electron and the hole $\rho_{eh}$ and their coordinates in the growth direction $z_e$ and $z_h$. The complete envelope wave function has the form
\begin{align}
    \psi^{{\bf Q}_X}({\bf r}_e, {\bf r}_h) = \frac{1}{\sqrt{A}} e^{i{\bf Q}_X\cdot{\bf R}^{\perp}_{eh}} \cdot \phi(\rho_{eh}, z_e, z_h),
\end{align}
where the exponential factor $e^{i{\bf Q}_X\cdot{\bf R}^\perp_{eh}}$ describes the in-plane free motion of the exciton as a whole. It is characterised by the in-plane 2D wave vector ${\bf Q}_X$ and depends on the 2D in-plane centre of mass vector ${\bf R}^{\perp}_{eh} = (m_e {\bf r}_e^{\perp} + m_h {\bf r}_h^{\perp})/M$, where $m_e$ and $m_{h}^{\perp}$ are the electron and hole effective masses in the plane of the QW, and $M = m_e + m_h^{\perp}$. Finally, $A$ is the normalization area.

For the carrier envelope wave functions, we use the finite square well model as a reasonable approximation of an ideal QW. For electrons and holes, the Hamiltonians are, respectively,
\begin{align}
    \hat{H}_e &= \frac{\hbar^2 \hat{k}^2_e}{2m_e} + V_e(z_e),\\
    \hat{H}_h &= \frac{\hbar^2}{2m_{h}^{\perp}} \left( \hat{k}^2_{h,x} + \hat{k}^2_{h,y} \right) + \frac{\hbar^2 \hat{k}^2_{h,z}}{2m_{h}^{_\parallel}} + V_h(z_h).
\end{align}
The total envelope wave function of a free electron (hole) $\psi^{{\bf Q}_{e(h)}}_{e(h)}({\bf r}_{e(h)})$ may be written as 
\begin{align}
    \psi_{e(h)}^{{\bf Q}_{e(h)}}({\bf r}_{e(h)}) &= \frac{1}{\sqrt{A}} e^{i{\bf Q}_{e(h)}\cdot{\bf r}_{e(h)}^{\perp}} \cdot \phi_{e(h)}(z_{e(h)}),
\end{align}
where $\phi_{e(h)}(z_{e(h)})$ is the wave function of the first quantum-confined state of the electron (hole), which depends on the carrier's coordinate in the growth direction $z_{e(h)}$. These wave functions have the well-known ``exp-cos-exp'' form (see, e.g.,~\cite{davies1997book}). The exponential factor describes the in-plane free motion of the particles, and is characterised by the carrier's in-plane 2D wave vector ${\bf Q}_{e(h)}$ and depends on it's coordinate in the QW plane ${\bf r}_{e(h)}^\perp$. $A$ is the same normalization area as before.

Apart from the spatial components of the exciton and carrier wave functions, we consider also the spin components. Let the growth axis be the quantisation axis for the angular momentum. The conduction band is assumed to be isotropic, and is characterised by just two spin projections, $s_e = \pm 1/2$. Neglecting hole subband mixing, two projections of angular momentum may be ascribed to the heavy-hole band, $j_h = \pm 3/2$. Therefore, a heavy-hole exciton has four independent states: the dipole-active (bright) excitons with total angular momentum $J_z = \pm 1$, and the dipole-forbidden (dark) excitons with $J_z = \pm 2$.

As in Ref.~\cite{ciuti1998}, we define the spin wave functions of such excitons in spin state $S_X$ as $\ket{S_X} = \chi^S(s_e, j_h)$. For example, $\ket{+1} = \chi^{+1}(s_e, j_h) = \delta_{s_e, -1/2}\delta_{j_h, +3/2}$  and $\ket{+2} = \delta_{s_e, +1/2}\delta_{j_h, +3/2}$, where $\delta$ is the Kronecker delta. In the general case of elliptically polarised light, excitons are created in the coherent spin superposition 
\begin{align} \label{E_alpha}
    \ket{E_\alpha} = \sin\alpha\ket{+1} + e^{i\phi}\cos\alpha\ket{-1}. 
\end{align}
The orthogonal state is $\ket{E_{\alpha+\pi/2}} = \cos\alpha\ket{+1} - \sin\alpha\ket{-1}$. Circular polarisation of excitation corresponds to $\alpha = 0, \pi/2$, while linear polarisation corresponds to $\alpha = \pi/4, 3\pi/4$. We note that the phase $\phi$ has a negligible effect on the following calculations, and so we will assume $\phi = 0$ for simplicity.

For electrons we define the spin wave functions $\ket{S_e} = \chi^{S_e}(s_e) = \delta_{s_e, S_e}$, and similarly for holes. The considered basis states for carriers are the spin states corresponding to the $z$-projection of spin:
\begin{align}
    \ket{S_e^\pm} &= \ket{\pm1/2}, \\
    \ket{S_h^\pm} &= \ket{\pm3/2}.
\end{align}
The total wave function of an exciton or carrier is taken as the product of the spatial and spin wave functions, i.e.
\begin{align}
    \Psi_{{\bf Q}_X}^{S_X}(e,h) = \psi^{{\bf Q}_X}({\bf r}_e, {\bf r}_h) \cdot \chi_{S_X}(s_e, j_h)
\end{align}
for excitons, and similarly for electrons and holes. Here the symbol $e$ ($h$) denotes both the spatial and spin components of the electron (hole).

\subsection{Scattering amplitudes}
\label{subsec:scatteringAmplitudes}

In this section, we will derive the scattering amplitudes for elastic scattering of $1s$-like excitons by $1s$-like excitons (X-X), and also by free electrons (X-e) and holes (X-h) in their respective ground state subbands. We consider the following scattering channels:
\newcommand*{\Scale}[2][4]{\scalebox{#1}{$#2$}}%
\newcommand*{\Resize}[2]{\resizebox{#1}{!}{$#2$}}%
\begin{align} \label{eqn:scattering_channels}
     \Scale[0.9]{{\rm X}} & \Scale[0.9]{{\rm -X}: ({\bf Q}_X,S_X) + ({\bf Q}_{X'}, S_{X'}) \rightarrow ({\bf Q}_{X}^f, S_{X}^f) + ({\bf Q}_{X'}^{f}, S_{X'}^{f}), }\nonumber\\
     \Scale[0.9]{{\rm X}} & \Scale[0.9]{{\rm -e}\ : ({\bf Q}_X,S_X) + ({\bf Q}_{e'}, S_{e'}) \ \rightarrow ({\bf Q}_{X}^{f}, S_{X}^{f}) + ({\bf Q}_{e'}^{f}, S_{e'}^{f}), }\nonumber\\
     \Scale[0.9]{{\rm X}} & \Scale[0.9]{{\rm -h}\:\:\!: ({\bf Q}_X,S_X) + ({\bf Q}_{h'}, S_{h'}) \:\!\rightarrow ({\bf Q}_{X}^{f}, S_{X}^{f}) + ({\bf Q}_{h'}^f, S_{h'}^{f}). }
\end{align}

\begin{figure}[h]
  \begin{center}
    \includegraphics[width=\linewidth]{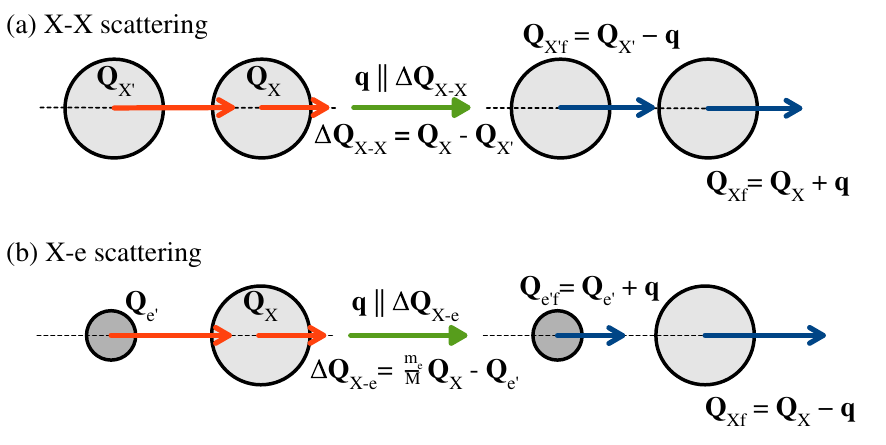}
  \end{center}
  \vspace{-.7cm}
  \caption{The X-X and X-e scattering processes and the definitions of the final wave vectors. For simplicity, all wave vectors are directed along the same line. The quantities $\Delta {\bf Q}_{X{\text - }e}$, $\Delta {\bf Q}_{X{\text -}X}$ are defined in Eqs.~(\ref{eqn:Xe_DeltaQ}), (\ref{eqn:XX_DeltaQ}). \iffalse We remark that the final wave vectors' magnitudes are not arbitrary, see the end of Sec.~\ref{sec:matrix_elements_calculation} \fi}
  \label{fig:scattering_expl}
\end{figure}

Here $({\bf Q}_X,S_X)$ is the $1s$-like exciton state with in-plane wave vector ${\bf Q}_X$ and spin $S_X$, and $({\bf Q}_{e'(h')}, S_{e'(h')})$ is the electron (hole) ground confined state with in-plane wave vector ${\bf Q}_{e'(h')}$ and spin $S_{e'(h')}$. In the above expressions, the scattering is elastic in the sense that the quasiparticles remain in their initial subbands after scattering. Because the total momentum must be conserved, we may rewrite the scattering channels in terms of the transferred momentum ${\bf q}$:
\begin{align} \label{eqn:transferred_momentum_def}
     \Scale[0.9]{{\rm X}} & \Scale[0.9]{{\rm -X}: {\bf Q}_{X}^{f} = {\bf Q}_X + {\bf q}; \qquad{\bf Q}_{X'}^{f} = {\bf Q}_{X'} - {\bf q}, }\nonumber\\
     \Scale[0.9]{{\rm X}} & \Scale[0.9]{{\rm -e}\ : {\bf Q}_{X}^{f} = {\bf Q}_X + {\bf q}; \qquad{\bf Q}_{e'}^{f} = {\bf Q}_{e'} - {\bf q}, }\\
     \Scale[0.9]{{\rm X}} & \Scale[0.9]{{\rm -h}\:\:\!: {\bf Q}_{X}^{f} = {\bf Q}_X + {\bf q}; \qquad{\bf Q}_{h'}^{f} = {\bf Q}_{h'} - {\bf q}. }\nonumber
\end{align}
An illustration of the scattering processes in the X-X and X-e cases is shown in Fig.~\ref{fig:scattering_expl}.

It is enough to consider only the channels (\ref{eqn:scattering_channels}) when the energy splittings between the ground and nearest excited subbands of the quasiparticles are greater than the calculated broadening and the average kinetic energies of the particles. At liquid helium temperatures the kinetic energies of thermalised quasiparticles are in the order of $0.5$~meV. Then, assuming the broadening to be in the order of 1~meV (as we will actually find), we must limit our treatment to QWs narrower than approximately 50~nm. In such QWs, the heavy-hole/light-hole splitting is larger than 1~meV, which allows us to neglect the scattering events in which a heavy-hole exciton transforms into a light-hole exciton. Conveniently, this also allows us to neglect the scattering of heavy-hole excitons by light-hole excitons, since in equilibrium at low temperatures they are practically absent from the sample, and quickly become heavy-hole excitons if created with resonant excitation. Another possible scattering channel is formed by transitions of heavy-hole excitons into their first excited $2s$-like state, however for a strictly 3D hydrogen-like exciton, the binding energy of the first excited state ($2s$) is only $1/4$ of that of the ground state. In GaAs QWs, this implies an energy splitting greater than $3$~meV, which effectively forbids such scattering events. Transitions with electrons/holes being excited into their second quantum-confined subbands may also be neglected, as the corresponding energy splittings are greater than those for excitons. 



We note that in this and the following sections, expressions for the exciton-carrier scattering will mostly only be written for electrons, as they are easily modified for the exciton-hole case.

In the  two-particle Hartree-Fock approximation, the properly symmetrised wave function of an exciton-electron system has the form
\begin{align} \label{eqn:phi_e(h)}
    \Phi_{{\bf Q}_X, {\bf Q}_{e'}}^{S, S_{e'}} = \frac{1}{\sqrt{2}} \bigl[ \Psi_{{\bf Q}_X}^{S_X}(e,h) \!&\cdot\! \Psi_{{\bf Q}_{e'}}^{S_{e'}}(e')   \nonumber\\
    - \Psi_{{\bf Q}_X}^{S_X}(e',h) \!&\cdot\! \Psi_{{\bf Q}_{e'}}^{S_{e'}}(e) \bigr].
\end{align}


Similarly, the wave function of the two-exciton system is defined as
\begin{align} \label{eqn:phi_X}
    \Phi_{{\bf Q}_X, {\bf Q}_{X'}}^{S_X, S_{X'}} = \frac{1}{2} \biggl\{\bigl[ &\Psi_{{\bf Q}_X}^{S_X}(e,h) \cdot \Psi_{{\bf Q}_{X'}}^{S_{X'}}(e',h')  \nonumber\\
    + &\Psi_{{\bf Q}_{X}}^{S_{X}}(e',h')  \cdot \Psi_{{\bf Q}_{X'}}^{S_{X'}}(e,h)  \bigr] \nonumber\\
    - [&\Psi_{{\bf Q}_{X}}^{S_{X}}(e',h)  \cdot \Psi_{{\bf Q}_{X'}}^{S_{X'}}(e,h')   \nonumber\\
    + &\Psi_{{\bf Q}_{X}}^{S_{X}}(e,h')  \cdot \Psi_{{\bf Q}_{X'}}^{S_{X'}}(e',h)  \bigr]  \biggr\}.
\end{align}
Here we must note that the wave functions (\ref{eqn:phi_e(h)}) and (\ref{eqn:phi_X}) form overcomplete non-orthogonal bases. This matter was studied by M. Combescot {\it et. al} and is dealt with in the framework of their ``coboson'' theory~\cite{combescot2008}. We, however, do not account for this non-orthogonality for the sake of simplicity. In Ref.~\cite{schindler2008} it has been pointed that in such an approach, a normalising denominator is left out. However, our estimates suggest that only a $\sim\!\!1 \%$ error in normalisation is induced by this approximation. We believe that these simplifications do not lead to significant errors in our calculations of the broadening. We also note that electron-hole exchange effects are neglected in this model. The corresponding energy splitting between the bright and dark exciton states reaches around 200~$\mu$eV in a 5~nm QW and quickly decreases in wider QWs~\cite{blackwood1994_e-h_exchange}. In this work, we neglect this splitting for simplicity.

The Hamiltonian of the exciton-electron system contains the kinetic energies of the three particles and all possible Coulomb interactions:
\begin{align}
    \hat{H}_{X{\text-}e} = \hat{T}_{X{\text-}e} &- V(|{\bf r}_e - {\bf r}_h|) \nonumber\\
    &+ V(|{\bf r}_e - {\bf r}_{e'}|) - V(|{\bf r}_{e'} - {\bf r}_h|).
\end{align}
In the two-exciton case:
\begin{align}
    \hat{H}_{X{\text-}X} = \hat{T}_{X{\text-}X} &- V(|{\bf r}_e - {\bf r}_h|) - V(|{\bf r}_{e'} - {\bf r}_{h'}|) \nonumber\\
    &+ V(|{\bf r}_e - {\bf r}_{e'}|) + V(|{\bf r}_h - {\bf r}_{h'}|) \nonumber\\
    &- V(|{\bf r}_e - {\bf r}_{h'}|) - V(|{\bf r}_{e'} - {\bf r}_h|).
\end{align}
In these expressions $\hat{T}_{X{\text -}X}$ and $\hat{T}_{X{\text -}e}$ contain the necessary kinetic energy operators, and $V(r) = e^2 / (\epsilon_0 r)$ is the Coulomb potential. The scattering amplitudes corresponding to the channels (\ref{eqn:scattering_channels}) have the basic form~\cite{ciuti1998}
\begin{align} \label{eqn:scattering_amplitude}
    H_{S_1 S_2}^{S_{1}^{f} S_{2}^{f}}({\bf Q}_1, {\bf Q}_2, {\bf q}) = \bra{\Phi_{{\bf Q}_1 + {\bf q}, {\bf Q}_2 - {\bf q}}^{S_{1}^{f} S_{2}^{f}}}\hat{H}\ket{\Phi_{{\bf Q}_1, {\bf Q}_2}^{S_1, S_2}},
\end{align}
where ${\bf Q}_1$ and $S_1$ describe the exciton state, while ${\bf Q}_2$ and $S_2$ stand for the state of the carrier ${\bf Q}_{e'(h')}$, $S_{e'(h')}$ or the state of the other exciton ${\bf Q}_{X'}$, $S_{X'}$. 

Let us first consider the exciton-electron case. The scattering amplitudes (\ref{eqn:scattering_amplitude}) are split into two terms:
\begin{align} \label{eqn:X-e_terms}
    \!\!\!H_{S_X S_{e'}}^{S_{X}^{f} S_{e'}^{f}} &({\bf Q}_X, \!{\bf Q}_{e'}, \!{\bf q}) \!=\! \braket{S_X|S_X^f}\!\braket{S_{e'}|S_{e'}^{f}}\! H_{\rm dir}^{X{\text -}e}({\bf Q}_X, \!{\bf Q}_{e'}, \!{\bf q}) \nonumber\\
     + &S_{\rm exch}^{X{\text -}e}(S_X, S_{e'}, S_X^f, S_{e'}^{f}) H_{\rm exch}^{X{\text -}e}({\bf Q}_X, \!{\bf Q}_{e'}, \!{\bf q}).
\end{align}
The first term ($H_{\rm dir}^{X{\text -}e}$) corresponds to the classical electrostatic interaction between the exciton and electron, and the second one ($H_{\rm exch}^{X{\text -}e}$) arises from an exchange of electrons. The spin factor is the spin exchange sum
\begin{align} \label{eqn:s^eh_exch}
    S_{\rm exch}^{X{\text -}e}(S_X, S_{e'}, S_{X}^f, S_{e'}^{f}) &= \sum_{s_e, j_h, s_{e'}} \chi^*_S(s_e, j_h) \chi^*_{S_{e'}}(s_{e'}) \nonumber\\
    &\times\chi_{S_f}(s_{e'}, j_h) \chi_{S_{e'}^{f}}(s_e).
\end{align}

As excitons are in the general case created by elliptically polarised light, on a short timescale the proper basis to consider for excitons is $(\ket{E_\alpha}, \ket{E_{\alpha+ \pi/2}}, \ket{\pm2})$ (see Eq.~(\ref{E_alpha})). For carriers, it is more practical to deal with the bases $\ket{\pm 1/2}$ and $\ket{\pm 3/2}$. The spin factors are easily calculated and are functions of $\alpha$. For a list of allowed exciton-carrier spin-scattering channels, see Table~\ref{table:S_exch^Xeh}.

\begin{table}[t]
\caption{\label{table:S_exch^Xeh}%
Non-zero spin exhange matrix elements for exciton-electron scattering. The considered bases are $\ket{\pm 1/2}$ for electrons and $(\ket{E_\alpha}, \ket{E_{\alpha+ \pi/2}}, \ket{\pm2})$ for excitons (see Sec. \ref{sec:theoretical}). For channels not indicated in the table, the factors can be obtained by using (\ref{eqn:s^eh_exch}). Similar results are easily obtained for holes with the basis $\ket{\pm 3/2}$.}
\begin{ruledtabular} {\renewcommand{\arraystretch}{1.5}
\begin{tabular}{cccc @{\hspace{0.5cm}}|c}
$S_X$ & $S_{e'}$ & $S_X^f$ & $S_{e'}^f$  & $S_{\rm exch}^{X{\text-}e}(S_X, S_{e'}, S_X^f, S_{e'}^{f})$ \\
\colrule
$E_\alpha$    &    $   +1/2$    &    $E_\alpha$                &   $ +1/2$    &    $\cos^2\alpha$\\
	$E_\alpha$    &    $    -1/2$    &    $E_\alpha$                &   $  -1/2$    &    $\sin^2\alpha$\\ 
	$E_\alpha$    &    $\pm1/2$    &    $E_{\alpha+\pi/2}$  & $\pm1/2$    &    $\rotpm\frac{1}{2}\sin2\alpha$\\ 
	$E_\alpha$    &    $    +1/2$   &   $   +2     $                 &   $  -1/2$    &     $\sin\alpha$\\ 
	$E_\alpha$    &    $    -1/2$    &    $   -2     $                  &   $  +1/2$    &    $\cos\alpha$\\
	\hline

	$  \pm  2 $     &    $\pm1/2$    &   $  \pm  2 $                 &    $\pm1/2$  &    $1$ \\
\end{tabular} }
\end{ruledtabular}
\end{table}

The direct Coulomb and electron exchange terms are the 9-dimensional integrals
 \begin{align}
    H_{\rm dir}^{X{\text -}e}&({\bf Q}_X, {\bf Q}_{e'}, {\bf q}) = \int {\rm d}^9 {\Omega}\  {\bf r}_{e'} \psi^*_{{\bf Q}_X}({\bf r}_e, {\bf r}_h) \psi^*_{{\bf Q}_{e'}}({\bf r}_{e'})  \nonumber\\ 
    \times & V_{X{\text-}e}({\bf r}_e, {\bf r}_h, {\bf r}_{e'})  \psi_{{\bf Q}_X + {\bf q}}({\bf r}_e, {\bf r}_h) \psi_{{\bf Q}_{e'}- {\bf q}}({\bf r}_{e'}), \label{eqn:H_dir^xe}\\
    H_{\rm exch}^{X{\text -}e}&({\bf Q}_X, {\bf Q}_{e'}, {\bf q}) =  \int {\rm d}^9 {\Omega}\  \psi^*_{{\bf Q}_X}({\bf r}_e, {\bf r}_h) \psi^*_{{\bf Q}_{e'}}({\bf r}_{e'})   \nonumber\\ 
    \times & V_{X{\text-}e}({\bf r}_e, {\bf r}_h, {\bf r}_{e'}) \psi_{{\bf Q}_X + {\bf q}}({\bf r}_{e'}, {\bf r}_h) \psi_{{\bf Q}_{e'} - {\bf q}}({\bf r}_e), \label{eqn:H_exch^xe} 
 \end{align}
where ${\rm d}^9 \Omega \eqDef {\rm d}^3 {\bf r}_e {\rm d}^3 {\bf r}_h  {\rm d}^3 {\bf r}_{e'}$, and $V_{X{\text-}e}$ is the interaction between the exciton and the electron:
\begin{align} \label{eqn:X-e_pot}
   \!\!\! V_{X{\text-}e}({\bf r}_e, {\bf r}_h, {\bf r}_{e'}) = V(|{\bf r}_e \!-\! {\bf r}_{e'}|) \!-\! V(|{\bf r}_h \!-\! {\bf r}_{e'}|).
\end{align}

Considering the four exponential factors in Eq.~(\ref{eqn:H_dir^xe}), we obtain a useful relation for the X-e fermion exchange integral $H_{\rm exch}^{X{\text -}e}$:

\begin{align} \label{eqn:Xe_DeltaQ}
    H_{\rm exch}^{X{\text -}e}({\bf Q}_X, {\bf Q}_{e'}, {\bf q}) &= H_{\rm exch}^{X{\text -}e}((m_e/M){\bf Q}_X - {\bf Q}_{e'}, {\bf q}) \nonumber\\
    &\eqDef H_{\rm exch}^{X{\text -}e}(\Delta{\bf Q}_{X{\text -}e}, {\bf q}).
\end{align}
A similar expression is true in the X-h case, with $\Delta{\bf Q}_{X{\text -}h} = (m_h/M){\bf Q}_X - {\bf Q}_{h'}$.

We will now consider exciton-exciton scattering. In this case we have four contributions to the scattering amplitude: 
\begin{align} \label{eqn:X-X_terms}
    &H_{S_X S_{X'}}^{S_{X}^{f} S_{X'}^f}({\bf Q}_X, {\bf Q}_{X'}, {\bf q}) \nonumber\\
     &= \braket{S_X |S_{X}^{f}}\braket{S_{X'}|S_{X'}^{f}} H_{\rm dir}({\bf Q}_X, {\bf Q}_{X'}, {\bf q}) \nonumber\\
     & + \braket{S_X|S_{X'}^{f}}\braket{S_{X'}|S_{X}^{f}} H_{\rm exch}^X({\bf Q}_X, {\bf Q}_{X'}, {\bf q}) \nonumber\\
     & + S_{\rm exch}^e(S_X, S_{X'}, S_{X}^{f}, S_{X'}^{f}) H_{\rm exch}^e({\bf Q}_X, {\bf Q}_{X'}, {\bf q}) \nonumber\\
     & + S_{\rm exch}^h(S_X, S_{X'}, S_{X}^{f}, S_{X'}^{f}) H_{\rm exch}^h({\bf Q}_X, {\bf Q}_{X'}, {\bf q}).
\end{align}


\begin{table*}[t]
\caption{\label{table:S_exch^eh} Allowed spin channels for exciton-exciton scattering and their respective spin exchange factors as reported in~\cite{ciuti1998}. The considered basis is $(\ket{E_\alpha}, \ket{E_{\alpha+ \pi/2}}, \ket{\pm2})$, where $\ket{E_\alpha}$ and $\ket{E_{\alpha+\pi/2}}$ are a pair of coherent orthogonal exciton spin states. For channels not indicated in the table, the factors can be obtained by using (\ref{eqn:s^X_exch}).}
\begin{ruledtabular} {\renewcommand{\arraystretch}{1.25}
\begin{tabular}{@{\hspace{1cm}} cccc @{\hspace{.75cm}}|cc @{\hspace{1cm}}}
$S_X$ & $S_{X'}$ & $S_X^f$ & $S_{X'}^f$  & $S_{\rm exch}^{e}(S_X, S_{X'}, S_X^f, S_{X'}^f)$ & $S_{\rm exch}^{h}(S_X, S_{X'}, S_X^f, S_{X'}^f)$\\ 
\colrule
$E_\alpha$    &    $E_\alpha$    &    $E_\alpha$    &   $E_\alpha$    &    $\sin^4\alpha + \cos^4\alpha$ &    $\sin^4\alpha + \cos^4\alpha$ \\
	$E_\alpha$    &    $E_\alpha$    &    $E_{\alpha+\pi/2}$   &   $E_{\alpha+\pi/2}$    &    $\frac{1}{2}\sin^2 2\alpha$ & $\frac{1}{2}\sin^2 2\alpha$\\ 
	$E_\alpha$    &    $E_\alpha$    &    $E_\alpha$   &   $E_{\alpha+\pi/2}$    &    $-\frac{1}{4}\sin4\alpha$ & $-\frac{1}{4}\sin4\alpha$\\ 
    $E_\alpha$    &    $E_\alpha$    &    $+2$   &   $-2$    &    $\frac{1}{2}\sin2\alpha$ & $\frac{1}{2}\sin2\alpha$\\ 
    \hline
	$E_\alpha$    &    $E_{\alpha+\pi/2}$    &    $E_\alpha$   &   $E_{\alpha+\pi/2}$    &    $\frac{1}{2}\sin^2 2\alpha$ & $\frac{1}{2}\sin^2 2\alpha$\\ 
	$E_\alpha$    &    $E_{\alpha+\pi/2}$    &    $E_{\alpha+\pi/2}$   &   $E_{\alpha+\pi/2}$    &    $\frac{1}{4}\sin4\alpha$ & $\frac{1}{4}\sin4\alpha$\\ 
	$E_\alpha$    &    $E_{\alpha+\pi/2}$    &    $+2$   &   $-2$    &    $-\sin^2\alpha$ & $\cos^2\alpha$\\
	 \hline 
    $E_\alpha$    &    $+2$    &   $E_\alpha$    &   $+2$    &    $\cos^2\alpha$ & $\sin^2\alpha$\\ 
    $E_\alpha$    &     $+2$   &    $E_{\alpha+\pi/2}$   &   $+2$    &    -$\frac{1}{2}\sin2\alpha$ & $\frac{1}{2}\sin2\alpha$\\ 
    \hline
    $E_\alpha$    &    $-2$    &   $E_\alpha$    &   $-2$    &    $\sin^2\alpha$ & $\cos^2\alpha$\\ 
    $E_\alpha$    &     $-2$   &    $E_{\alpha+\pi/2}$   &   $-2$    &    $\frac{1}{2}\sin2\alpha$ & $-\frac{1}{2}\sin2\alpha$\\
	\hline
	$\pm 2$ & $\pm 2$ & $\pm 2$ & $\pm 2$ & 1 & 1\\
\end{tabular} }
\end{ruledtabular}
\end{table*}


These contributions correspond respectively to the classic electrostatic interaction between the two excitons ($H_{\rm dir}$), exciton-exciton exchange ($H_{\rm exch}$), and fermion exchange related to the exchange of only electrons and holes ($H_{\rm exch}^e$, $H_{\rm exch}^h$), respectively. The spin factors are the spin-exchange sums
\begin{align} \label{eqn:s^X_exch}
    S_{\rm exch}^{e(h)}(S_X, &S_{X'}, S_X^f, S_{X'}^f) = \sum\limits_{\substack{s_e, j_h, \\s_{e'}, j_{h'}}} \chi_S^*(s_e, j_h) \chi_{S'}^*(s_{e'}, j_{h'})  \nonumber\\ 
    &\times \chi_{S_f}(s_{e'(e)}, j_{h(h')}) \chi_{S'_{f}}(s_{e(e')}, j_{h'(h)}).
\end{align}
We remind that $(\ket{E_\alpha}, \ket{E_{\alpha+ \pi/2}}, \ket{\pm2})$ is the appropriate spin-state basis to consider for excitons. The allowed exciton-exciton spin scattering channels as reported in~\cite{ciuti1998} are listed in Table~\ref{table:S_exch^eh}.

The direct Coulomb and the three exchange integrals are the 12-dimensional integrals
\begin{align} \nonumber
    &H_{\rm dir}({\bf Q}_{X}, \!{\bf Q}_{X'}, \!{\bf q}) = \int {\rm d}^{12}\Omega\  {\bf r}_{h'} \psi_{{\bf Q}_X}^*({\bf r}_e, {\bf r}_h) \psi_{{\bf Q}_{X'}}^*({\bf r}_{e'}, {\bf r}_{h'}) \label{eqn:H_dir}\\
    & \times V_{X{\text-}X}({\bf r}_e, {\bf r}_h, {\bf r}_{e'}, {\bf r}_{h'}) \psi_{{\bf Q}_X + {\bf q}}({\bf r}_e, {\bf r}_h) \psi_{{\bf Q}_{X'} - {\bf q}}({\bf r}_{e'}, {\bf r}_{h'}), \\ \nonumber 
    &H_{\rm exch}^X({\bf Q}_{X}, {\bf Q}_{X'}, {\bf q}) = \int {\rm d}^{12}\Omega\  \psi_{{\bf Q}_X}^*({\bf r}_e, {\bf r}_h) \psi_{{\bf Q}_{X'}}^*({\bf r}_{e'}, {\bf r}_{h'}) \label{eqn:H_exch^X}\\
    & \times V_{X{\text-}X}({\bf r}_e, {\bf r}_h, {\bf r}_{e'}, {\bf r}_{h'}) \psi_{{\bf Q}_X + {\bf q}}({\bf r}_{e'}, {\bf r}_{h'}) \psi_{{\bf Q}_{X'} - {\bf q}}({\bf r}_e, {\bf r}_h), \\ \nonumber
    &H_{\rm exch}^e({\bf Q}_{X}, {\bf Q}_{X'}, {\bf q}) = \int {\rm d}^{12}\Omega\  \psi_{{\bf Q}_X}^*({\bf r}_e, {\bf r}_h) \psi_{{\bf Q}_{X'}}^*({\bf r}_{e'}, {\bf r}_{h'}) \label{eqn:H_exch^e}\\
    & \times V_{X{\text-}X}({\bf r}_e, {\bf r}_h, {\bf r}_{e'}, {\bf r}_{h'}) \psi_{{\bf Q}_X + {\bf q}}({\bf r}_{e'}, {\bf r}_h) \psi_{{\bf Q}_{X'} - {\bf q}}({\bf r}_e, {\bf r}_{h'}), \\ \nonumber
    &H_{\rm exch}^h({\bf Q}_{X}, {\bf Q}_{X'}, {\bf q}) = \int {\rm d}^{12}\Omega\  \psi_{{\bf Q}_X}^*({\bf r}_e, {\bf r}_h) \psi_{{\bf Q}_{X'}}^*({\bf r}_{e'}, {\bf r}_{h'}) \label{eqn:H_exch^h}\\
    & \times V_{X{\text-}X}({\bf r}_e, {\bf r}_h, {\bf r}_{e'}, {\bf r}_{h'}) \psi_{{\bf Q}_X + {\bf q}}({\bf r}_e, {\bf r}_{h'}) \psi_{{\bf Q}_{X'} - {\bf q}}({\bf r}_{e'}, {\bf r}_h),
\end{align}
where ${\rm d}^{12} \Omega \eqDef {\rm d}^3 {\bf r}_e {\rm d}^3 {\bf r}_h  {\rm d}^3 {\bf r}_{e'} {\rm d}^3 {\bf r}_{h'}$ and $V_{X{\text-}X}$ is the interaction between the two excitons:
\begin{align} \label{eqn:X-X_pot}
    \!\!\!V_{X{\text-}X}({\bf r}_e, {\bf r}_h, {\bf r}_{e'}, {\bf r}_{h'}) \!&=\! V(|{\bf r}_e \!-\! {\bf r}_{e'}|) \!+\! V(|{\bf r}_h \!-\! {\bf r}_{h'}|) \nonumber\\
    &-\! V(|{\bf r}_e \!-\! {\bf r}_{h'}|) \!-\! V(|{\bf r}_{e'} \!-\! {\bf r}_h|).
\end{align}

One may notice that $H_{\rm dir}({\bf Q}_X, {\bf Q}_{X'}, {\bf q}) \equiv H_{\rm dir}({\bf q})$ (the exponents cancel each other out), and that $H_{\rm exch}^X({\bf Q}_X, {\bf Q}_{X'}, {\bf q}) = H_{\rm dir}({\bf Q}_X, {\bf Q}_{X'}, {\bf Q}_{X'} - {\bf Q}_X - {\bf q})$~\cite{ciuti1998}. Particularly, when ${\bf Q}_X = {\bf Q}_{X'} = 0$, 
\begin{align} \label{eqn:dir_exch_relation}
    H_{\rm dir}({\bf q}) = H_{\rm exch}^X({\bf q}) \eqDef H_{\rm dir}^{X{\text -}X}({\bf q}).
\end{align}
For the fermion exchange integrals it can be shown that~\cite{ciuti1998}
\begin{align} \label{eqn:XX_DeltaQ}
    H_{\rm exch}^{e(h)}({\bf Q}_X, {\bf Q}_{X'}, {\bf q}) &= H_{\rm exch}^{e(h)}({\bf Q}_X - {\bf Q}_{X'}, {\bf q}) \nonumber\\
    &\eqDef H_{\rm exch}^{e(h)}(\Delta{\bf Q}_{X{\text -}X}, {\bf q}).
\end{align}
Moreover, if $\Delta{\bf Q}_{X{\text -}X} = 0$, then 
\begin{align} \label{eqn:fermion_exch_q_relation}
    H_{\rm exch}^e({\bf q}) = H_{\rm exch}^h({\bf q}) \eqDef H_{\rm exch}^{X{\text -}X}({\bf q}).
\end{align}

We note that all of the matrix elements (\ref{eqn:H_dir^xe})-(\ref{eqn:H_exch^xe}) and (\ref{eqn:H_dir})-(\ref{eqn:H_exch^h}) are real because of symmetry considerations (see Ref.~\cite{ciuti1998} for details). Also, it is necessary to point out that all of the integrals above are inversely proportional to the normalization area $A$: the product of the wave functions is proportional to $1/A^2$, while integrating over the systems' centre of mass gives an $A$ factor. Physically, this means that the farther apart the excitons are from carriers or other excitons on average, the weaker the scattering. Of course, when calculating observables such as the collisional broadening, the end result does not depend on $A$. Still, it is easier to work with normalisation-independent quantities. Therefore we will be dealing with the quantities $J_{\rm dir}^{X{\text -}e(h)}$, $J_{\rm exch}^{X{\text -}e(h)}$, $J_{\rm dir}^{X{\text -}X}$, $J_{\rm exch}^{X{\text -}X}$, which are related to the matrix elements by the simple expression
\begin{align} \label{eqn:H_to_J}
    J = A\cdot H,
\end{align}
and are independent of $A$. In particular, at zero wave vectors ($\Delta{\bf Q}_{X{\text -}X} = \Delta{\bf Q}_{X{\text -}e(h)} = {\bf q} = 0$) the integrals are commonly known as the exchange constants (see, e.g.,~\cite{landafshitz} and \cite{trifonov2019PRL} and it's Supplemental Material). Their physical meaning is evident from the relation $A = 1/n_{X,e,h}$, where $n_{X,e,h}$ is the areal density of the respective quasiparticles.

\section{Calculation of matrix elements}
\label{sec:matrix_elements_calculation}

To calculate the wave functions for excitons and carriers we use material parameters used in~\cite{khramtsov2016}, apart from the effective masses, which were taken from the work by I. Vurgaftman {\it et. al}~\cite{Vurgaftman2001}.

The eigenfunctions of the discretised 3-dimensional Schr\"odinger equation are represented by an array of wave function amplitudes distributed across a uniform 3D-grid with step $h$. Generally, to obtain precise values for the matrix elements one must evaluate a series of them using wave functions with decreasing step $h$, the precise value being the limit of this dependence as $h\to0$ (see~\cite{khramtsov2016} for details). However, as an estimate, it is often sufficient to evaluate the matrix elements using one finite but sufficiently small $h$, as will be shown below.

Since we use numerical wave functions for the relative motion inside an exciton, all of the integrals must be computed numerically. For each set of momenta $\{\Delta{\bf Q}, {\bf q}\}$ the integrals in question are 9- and 12-dimensional, so even a single point of the momentum dependences presents a certain computational challenge. However, there are ways to drastically speed up the calculations. Firstly, a change of variables is necessary (see \hyperref[sec:appendixA]{Appendix A}). Then, by efficiently implementing a standard Monte-Carlo integration scheme utilising GPU parallelism with the help of NVidia CUDA, we were able to speed up the calculations by orders of magnitude. The calculation of a single exciton-exciton integral to an error of around $10\%$ takes several minutes (on an NVidia GTX1660 6GB GPU) at around $10^{9}-10^{10}$ 10-dimensional points, whereas it would take hours on a regular CPU. For carriers the calculations are much quicker \cite{source_code}. 
\begin{figure}[h]
  \begin{center}
    \includegraphics[]{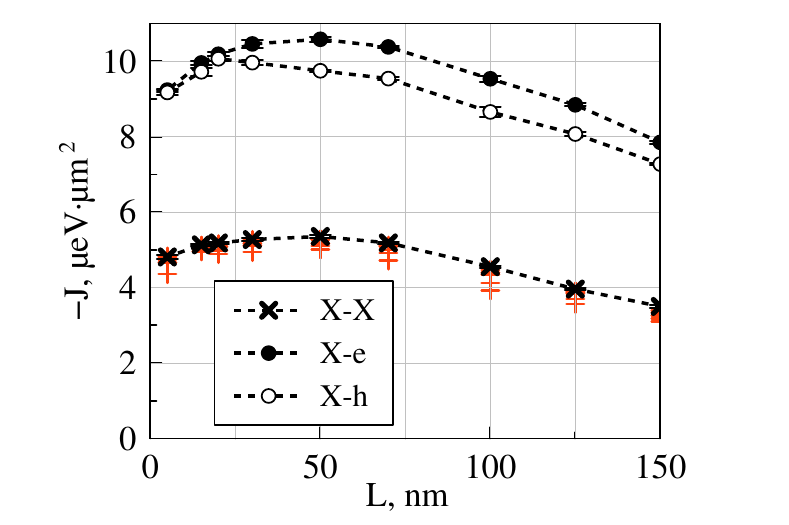}
  \end{center}
  \vspace{-.5cm}
  \caption{Fermion exchange constants for exciton-exciton (black crosses), exciton-electron (dots) and exciton-hole (empty dots) as functions of QW width $L$. Each point is extrapolated from sequences calculated at decreasing step values $h$. Red crosses denote examples of such sequences for the exciton-exciton curve. Note that there are two X-X fermion exchange terms, corresponding to electron-electron and hole-hole exchange, which are characterised by the same exchange constant, shown in the figure (see Eqs.~(\ref{eqn:X-X_terms}) and (\ref{eqn:fermion_exch_q_relation})). We neglect heavy-hole/light-hole mixing, therefore the results at $L>50$~nm should be treated as rough estimates. The general decrease of the exchange constants at large QW width is not a consequence of this.}
  \label{fig:J_all}
\end{figure}

The exchange constants for fermion exchange in the X-e, X-h and X-X cases as functions of QW width are plotted in Fig.~\ref{fig:J_all}. In this figure, each black symbol is an extrapolation of the dependences $J(L, h)$ as $h\to 0$. For the exciton-exciton curve the finite-$h$ points have been added for reference, shown by red crosses. The largest $h$ used in these extrapolations are in the range $1.25$--$4$~nm, depending on the QW width. As it can be seen from the figure, the extrapolation does not alter the values significantly; in fact the smallest $h$ ($0.25$--$1$~nm, depending on the QW width) generally give very good estimates, with errors in the order of 1\%. Nevertheless, the extrapolation must be performed in order to obtain a smooth curve. As we will see, the broadening's dependence on the QW width $L$ is mainly determined by the curves in Fig.~\ref{fig:J_all}, i.e. the exchange constants.
\begin{figure*}[ht!]
    \centering
    \includegraphics[]{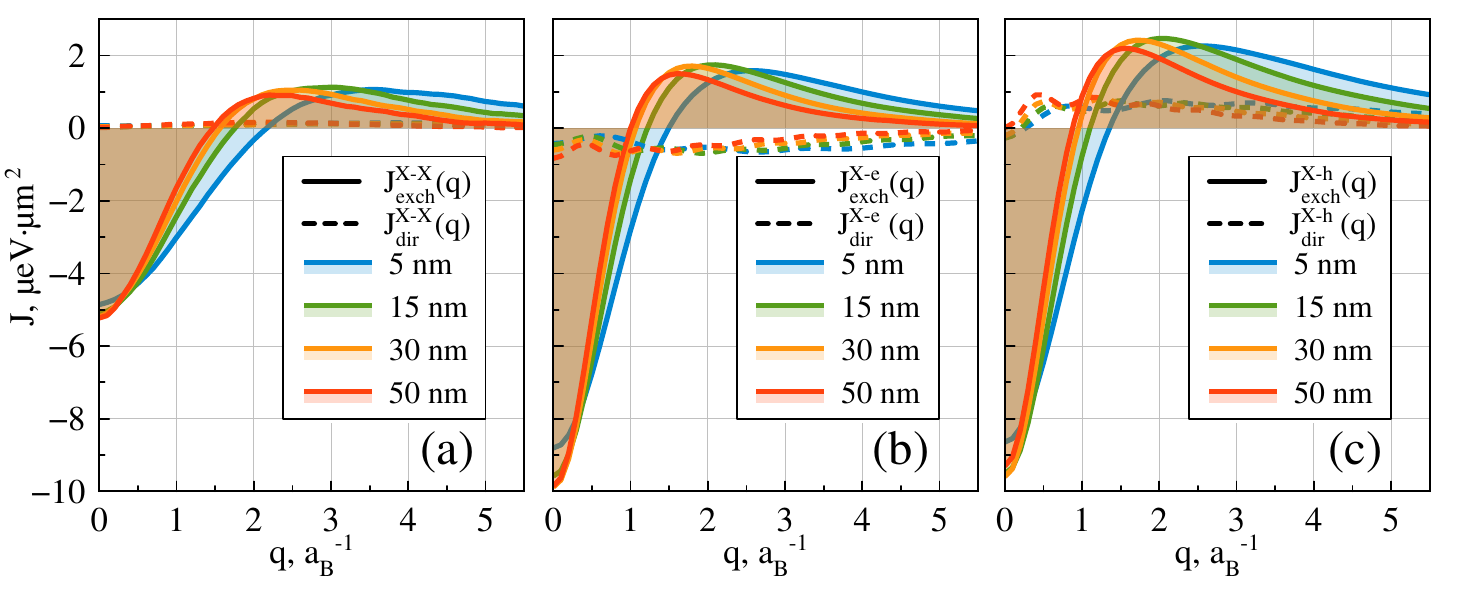}
    \caption{Matrix elements of exciton-exciton (a), exciton-electron (b), and exciton-hole (c) scattering as functions of transferred momentum $q$ at $\Delta{\bf Q}_{X{\text -}X} = \Delta{\bf Q}_{X{\text -}e(h)} = 0$. Dotted lines: direct terms, solid lines: exchange terms. Colours denote QW widths. The exciton wave function's grid step $h$ for each QW width used: 5~nm - 0.25~nm, 15~nm - 0.3~nm, 30~nm - 0.5~nm, 50~nm - 1~nm. $a_{\rm B} = 11.7$~nm is the bulk GaAs hydrogen-like heavy-hole exciton Bohr radius. Note that there are four terms in the X-X interaction represented by the two integrals in panel (a) (see Eqs.~(\ref{eqn:X-X_terms}) and (\ref{eqn:fermion_exch_q_relation})).}
    \label{fig:J_q_tryptich}
\end{figure*}

The negative signs of the exchange constants imply the tendency of antiparallel spin alignment. Noting that the exciton-exciton interaction includes two fermion exchange terms (see Eqs.~(\ref{eqn:X-X_terms}) and (\ref{eqn:fermion_exch_q_relation})), we can see that the exchange interaction is of similar magnitude in the three systems studied. The exciton-hole exchange constants are slightly smaller in magnitude than those in the exciton-electron case. This is a consequence of the hole's larger mass. In narrow wells the X-e and X-h curves coincide. In this region, the hole's larger mass leads to weaker barrier penetration, which in turn enhances the interaction relative to the exciton-electron case. Overall the QW width dependences of the exchange constants are rather weak, with variation of around 10\% in the considered range. The broad maxima of the dependences in the region of $L = 25$--$65$~nm are likely a result of the interplay between the increasing confinement and the changing shape of the wave functions, both along the growth axis, and in the $(x, y)$ plane. With increasing well width, the average distance between the particles along the growth axis also increases, and this leads to a decrease in the interactions' strength.

C. Ciuti {\it et. al} report the fermion exchange constant for exciton-exciton interaction to be equal to approximately 4.2~$\mu$eV$\cdot\mu$m$^2$ for strictly 2D excitons in GaAs~\cite{ciuti1998}, which is in good agreement with our data for narrow QWs. For an infinite-barrier 20~nm GaAs QW the exciton-electron exchange constant is around 15~$\mu$eV$\cdot\mu$m$^2$, as reported by G. Ramon, A. Mann and E. Cohen~\cite{cohen_neutral&charged_exciton-electron2003}, whereas our results estimate it to be around 10~$\mu$eV$\cdot\mu$m$^2$. In their work, the effective Bohr radius of a trial exciton wave function was calculated by a variational procedure and turned out to be around $15$~nm. According to our estimates, it is smaller by around 15\%. Although this difference may not affect single exciton integrals, the different behaviour of their approximate wave function and our numerical exciton wave function at large $\rho_{eh}$ could significantly alter the exciton-electron integrals.

Obtaining the matrix elements as proper functions of all the considered momenta (see Eq.~(\ref{eqn:scattering_amplitude})) involves calculating 7- and 10-dimensional integrals as functions of 3 scalar variables~\cite{ciuti1998, cohen_neutral&charged_exciton-electron2003}. Estimating around 10 points per variable, we would require $10^3$ integrals for each interaction type and each QW considered, which is an incredibly difficult computational task. Let us consider the simpler case of $\Delta{\bf Q}_{X{\text -}X} = \Delta{\bf Q}_{X{\text -}e(h)} = 0$, which corresponds to low temperatures and low pumping intensity, when the average momentum of all quasiparticles is small compared to the inverse exciton Bohr radius. The matrix elements calculated as functions of transferred momentum $q$ for several QW widths are shown in Fig.~\ref{fig:J_q_tryptich}. The chosen QWs well represent the region where our wave functions are applicable, that is, where the QWs are neither too narrow for the envelope function approximation to be accurate, nor too wide for the effects of heavy-hole/light-hole mixing to be significant. In these calculations we have not performed $h$-extrapolation, using exciton wave functions with relatively small steps as a reasonable approximation (see caption to Fig.~\ref{fig:J_q_tryptich}). 

It can clearly be seen that at all $L$ the fermion exchange terms dominate, which is a fact well established in literature~\cite{amand1997}; in fact, the overall shape of the $q$-dependence of the matrix elements is well-known~\cite{ciuti1998, cohen_neutral&charged_exciton-electron2003, ouerdane2008}. The specific shape of the curves in Fig.~\ref{fig:J_q_tryptich} is determined by the interplay of the attractive and repulsive potentials in Eqs.~(\ref{eqn:X-e_pot}) and (\ref{eqn:X-X_pot}) and the different confinement of the electron and hole inside an exciton. The matrix elements of single Coulomb potentials (the terms in Eqs.~(\ref{eqn:X-e_pot}) and (\ref{eqn:X-X_pot})) are approximately an order of magnitude higher than those of the total potentials $V_{X{\text -}X(e,h)}$. The total charge of an exciton is zero, therefore at $q=0$ the direct terms are negligible, vanishing completely for all $q$ if $m_e=m_h$ because of equal confinement of the Coulomb-bound electron and hole. The exchange integrals, in contrast, reach their maximum values at $q=0$, equal to the exchange constants shown in Fig.~\ref{fig:J_all}.

\section{Collisional broadening}
\label{sec:broadening}

In optical experiments, the scattering of quasiparticles in general leads to lineshifts and homogeneous collisional broadening. One possible approach to calculating this density-dependent broadening has been described in detail in~\cite{dense_gas_1} and~\cite{dense_gas_2}, and successfully used in~\cite{ciuti1998}. In this approach, based on the second-order Born approximation, an implicit equation for the broadening $\Gamma_{{\bf Q}_X}^{S_X}$ of the exciton state (${\bf Q}_X$, $S_X$) in the case of exciton-exciton scattering is derived:
\begin{align} \label{eqn:Gamma_XX_1}
    \!\!\!\!\!\Gamma_{{\bf Q}_X}^{S_X} \!&= 2\pi \!\sum_{{\bf Q}_{X'}} \!\!\sum\limits_{\substack{S_{X'},\\ S_X^f, S_{X'}^f}} \!\!\!\! N_{S_{X'}}({\bf Q}_{X'})
    \!\sum_{{\bf q} \neq 0}      \left|  H_{S_X S_{X'}}^{S_X^f S_{X'}^f} ({\bf Q}_X, \!{\bf Q}_{X'}, \!{\bf q}) \right|^2  \nonumber\\
    &\times \mathcal{L}(E_{{\bf Q}_X} + E_{{\bf Q}_{X'}} - E_{{\bf Q}_X + {\bf q}} - E_{{\bf Q}_{X'} - {\bf q}}, \nonumber\\
    &\qquad\qquad\Gamma_{{\bf Q}_X}^{S_X} + \Gamma_{{\bf Q}_{X'}}^{S_{X'}} + \Gamma_{{\bf Q}_X + {\bf q}}^{S_X^f} + \Gamma_{{\bf Q}_{X'} - {\bf q}}^{S_{X'}^f}).
\end{align}
Similar expressions may be derived for exciton-carrier scattering. For electrons
\begin{align} \label{eqn:Gamma_Xe_1}
    \!\!\!\!\!\Gamma_{{\bf Q}_X}^{S_X} \!&= 2\pi \sum_{{\bf Q}_{e'}} \!\!\sum\limits_{\substack{S_{e'},\\ S_X^f, S_{e'}^f}} N_{S_{e'}}({\bf Q}_{e'}) 
    \!\sum_{{\bf q} \neq 0}      \left|  H_{S_X S_{e'}}^{S_X^f S_{e'}^{f}} ({{\bf Q}_X}, \!{\bf Q}_{e'}, \!{\bf q}) \right|^2 \nonumber \\
    &\times \mathcal{L}(E_{{\bf Q}_X} + \mathcal{E}_{{\bf Q}_{e'}} - E_{{\bf Q}_X + {\bf q}} - \mathcal{E}_{{\bf Q}_{e'} - {\bf q}}, \nonumber\\
    &\qquad\qquad\Gamma_{{\bf Q}_X}^{S_X} + \Gamma_{{\bf Q}_{e'}}^{S_{e'}} + \Gamma_{{\bf Q}_X + {\bf q}}^{S_X^f} + \Gamma_{{\bf Q}_{e'} - {\bf q}}^{S_{e'}^{f}}).
\end{align}
In the equations above $N_{S_{X'}}({\bf Q}_{X'})$ ($N_{S_{e'}}({\bf Q}_{e'})$) is the number of excitons (electrons) in the corresponding state, $E_{\bf Q} = \hbar^2 Q^2 / 2 M$ and $\mathcal{E}_{{\bf Q}_{e'}} = \hbar^2 Q_{e'}^2 / 2 m_e$ are the exciton and electron in-plane kinetic energies respectively, and $\mathcal{L}$ is the Lorentzian function
\begin{align}
    \mathcal{L}(E, \gamma) = \frac{1}{\pi} \frac{\gamma/2}{E^2 + (\gamma/2)^2}.
\end{align}

The physical meaning of equations (\ref{eqn:Gamma_XX_1}) and (\ref{eqn:Gamma_Xe_1}) is clear: because of additional interactions, excitons are damped quasiparticles, and the conservation of energy is partially lifted\footnote{This may seem contradictory to us neglecting scattering to excited states, however, the obtained broadening is in the order of 1~meV. This is not enough for the possible transitions described in Sec. \ref{subsec:scatteringAmplitudes}}. The finite-width Lorentzian is a consequence of the two-exciton or exciton-carrier energy having an imaginary component, which itself arises from the non-hermiticity of the effective Hamiltonian~\cite{dense_gas_2}. In the limit of very small damping, i.e. as ${\gamma\to 0}$, $\mathcal{L}(E, \gamma)\to \delta(E)$, and the expressions (\ref{eqn:Gamma_XX_1}), (\ref{eqn:Gamma_Xe_1}) are reduced to the usual Born approximation (see, for example {\it Collision theory} in~\cite{gasiorowicz2007quantum}).

In considering these equations, we note that the functions $N_{S_{X'}}({\bf Q}_{X'})$ and $N_{S_{e'}}({\bf Q}_{e'})$ depend on the experimental conditions (i.e., temperature, pump energy and intensity, pump polarisation etc.). In the general case they must be obtained separately by considering the dynamics of the exciton (carrier) gas, which is beyond the scope of our work. However, the problem can be simplified under certain conditions.

We assume the excitons to be created by resonant photons, thus their wave vector is around 0.3 in units of $1/a_{\rm B}$ ($a_{\rm B} = 15.9$~nm is the bulk GaAs heavy-hole exciton Bohr radius). On a timescale of several ps the excitons remain largely within the light cone. At longer times, the excitons thermalise, and the non-radiative exciton reservoir is formed~\cite{kurdyubov2021arXiv}. At helium temperatures, the wave vectors of reservoir excitons ${\bf Q}$ reach values of about 0.5--0.7 in units of $1/a_{\rm B}$. We have carried out additional calculations with finite in-plane momenta in a 30~nm QW, and they reveal that at such wave vectors, the scattering matrix elements do decrease relative to the ${\bf Q}_X = 0$ case, but not significantly, similarly to the $q$-dependences shown in Fig.~\ref{fig:J_q_tryptich}. Thus, even the large wave vector exciton reservoir can be approximately treated as if ${\bf Q}_X = 0$ when discussing exciton-exciton and exciton-carrier interactions, provided it is characterised by temperatures of less than 10~K. Nevertheless, we will  mainly restrict our discussion to excitons inside the light cone, which are well-approximated by ${\bf Q}_X = 0$. 
 
Carriers may also be created optically. Hot carriers created by non-resonant excitation are characterised by large kinetic energies in the order of 10--30~meV, which implies electron wave vectors in the range 1.5--3.0 in units of $1/a_B$. For holes, the corresponding wave vectors are larger by a factor of around 1.3. In our calculations of the broadening below we neglect the carrier momenta ${\bf Q}_{e'(h')}$, which implies a low temperature carrier reservoir. This is possible with near-resonant excitation, and also in experiments with low intensity continuous-wave (CW) pumping. If discussing dynamics experiments, then only processes with characteristic times much greater than the carrier energy relaxation time should be considered. It is well known that the energy relaxation of hot carriers is a fast process, with characteristic times in the order of 100~ps (see, e.g., Ref.~\cite{balkan1989}).

To further simplify the problem, we will also neglect the dependence of the broadening $\Gamma$ on spin and momentum. Moreover, we will assume that in the exciton-carrier case the damping is the same for excitons and carriers. This is a fairly crude approximation, however the damping values should be in the same order for excitons and carriers, which is indeed the case, as will be shown below.

Finally, replacing the sum over ${\bf q}$ by an integral and rewriting in terms of $J = A\cdot H$ (see Eq.~(\ref{eqn:H_to_J})), we transform equations (\ref{eqn:Gamma_XX_1}) and (\ref{eqn:Gamma_Xe_1}) to
\begin{align}  \label{eqn:Gamma_XX_2}
    \Gamma_{X{\text-}X}  = \sum\limits_{\substack{S_{X'},\\ S_X^f, S_{X'}^f}} 
    \int_{0}^{\infty} {\rm d}q\   &q    \left|  J_{S_X S_{X'}}^{S_X^f S_{X'}^f} (0, 0, q) \right|^2 \nonumber\\[-.5cm]
    &\times \frac{1}{\pi} \frac{2\Gamma_{X{\text-}X}}{\left(\frac{\hbar^2 q^2}{M}\right)^2 + 4\Gamma_{X{\text-}X}^2},
\end{align}
\begin{align}  \label{eqn:Gamma_Xe_2}
    \Gamma_{X{\text-}e}  = \sum\limits_{\substack{S_{e'},\\ S_X^f, S_{e'}^f}} 
    \int_{0}^{\infty}{\rm d}q\   &q    \left|  J_{S_X S_{e'}}^{S_X^f S_{e'}^{f}} (0, 0, q) \right|^2 \nonumber\\[-.5cm]
    &\times \frac{1}{\pi} \frac{2\Gamma_{X{\text-}e}}{\left(\frac{\hbar^2 q^2}{2M^*_e}\right)^2 + 4\Gamma_{X{\text-}e}^2},
\end{align}
where we have introduced the exciton (electron) areal density $n_{S_{X'}} = N_{S_{X'}} / A$ ($n_{S'_e} = N_{S'_e} / A$) and have defined $M_e^* = 1/(M^{-1} + m_e^{-1})$. Expression (\ref{eqn:Gamma_Xe_2}) is trivially modified for holes.

To obtain the broadening, we must specify the distributions $n_{S}$, $n_{S_{e(h)}}$, which determine the spin polarisation of the ensemble of scattering particles. The parameter $\alpha$ reflects the polarisation of photons which create new excitons, and determines the spin scattering channels (see Tables \ref{table:S_exch^Xeh}, \ref{table:S_exch^eh}). We consider the exciton-exciton problem first. 
In the case of circularly polarised light ($\alpha = 0, \pi/2$), excitons are created in the states $\ket{\pm 1}$. If all excitons are created optically, then only the transitions $(\pm 1, \pm 1)\to(\pm 1, \pm 1)$ take place. This means that the exciton population remains in the initial state as long as other spin relaxation mechanisms are negligible.  Assuming a short-delay pump-probe experiment scheme with co-polarised pump and probe pulses, we then have
\begin{align} \label{eqn:Gamma_XX_f_circ}
    \Gamma_{X{\text-}X}  = n_X
    \int_{0}^{\infty}{\rm d}q\   &q    \left|  J_{1,1}^{1,1} (0, 0, q) \right|^2 \nonumber\\[-.0cm]
    &\times \frac{1}{\pi} \frac{2\Gamma_{X{\text-}X}}{\left(\frac{\hbar^2 q^2}{M}\right)^2 + 4\Gamma_{X{\text-}X}^2},
\end{align}
 which we will refer to as the co-polarised case.

Now let us assume an experiment with excitation into the heavy-hole exciton resonance by a linearly polarised laser pulse ($\alpha = \pi/4$). In this case two bright excitons in the state $\ket{E_\alpha}$ can scatter not only into the same states or orthogonal ones, but also to the dark states $\ket{\pm2}$. Considering all the possible spin channels and their relative weight (see Table \ref{table:S_exch^eh}), we find that eventually an equal redistribution of populations in each state must take place. The time required for this redistribution is in the order of the inverse of the broadening (several ps for a broadening of $\sim\!$ 1~meV). In this case $n_{\ket{x}} = n_{\ket{y}} = n_{\ket{+2}} = n_{\ket{-2}} = n_X / 4$ and we have
\begin{align}  \label{eqn:Gamma_XX_f_lin}
    \Gamma_{X{\text-}X}  = \sum\limits_{\substack{S_{X'},\\ S_X^f, S_{X'}^f}} \frac{n_X}{4} 
    \int_{0}^{\infty}&{\rm d}q\   q    \left|  J_{S_X S_{X'}}^{S_X^f S_{X'}^f} (0, 0, q) \right|^2 \nonumber\\[-.5cm]
    &\times \frac{1}{\pi} \frac{2\Gamma_{X{\text-}X}}{\left(\frac{\hbar^2 q^2}{M}\right)^2 + 4\Gamma_{X{\text-}X}^2}.
\end{align}
 This equation may be applied to luminescence experiments or experiments where a cold exciton reservoir is formed. We note that in the case of equal population of each state, the choice of $\alpha$ is arbitrary, as one would expect.

For the problem of exciton-carrier scattering similar reasoning may be employed. 
In the limiting case of complete polarisation of both excitons and carriers, only the transitions $(\ket{\pm 1}, \ket{\rotpm 1/2})\to(\ket{\pm 1}, \ket{\rotpm 1/2})$ take place (for holes it is the same channel with $\ket{\pm 3/2}$ instead of $\ket{\rotpm 1/2}$). Therefore  for a short-delay co-polarised pump-probe experiment we may write the exciton-electron collisional broadening as
\begin{align}   \label{eqn:Gamma_Xe_f_circ}
    \Gamma_{X{\text-}e}  = n_e
    \int_{0}^{\infty}{\rm d}q\   &q    \left|  J_{1,-1/2}^{1,-1/2} (0, 0, q) \right|^2  \nonumber\\[-.0cm]
    &\times \frac{1}{\pi} \frac{2\Gamma_{X{\text-}e}}{\left(\frac{\hbar^2 q^2}{2M^*_e}\right)^2 + 4\Gamma_{X{\text-}e}^2},
\end{align}
and for holes a similar expression is obtainable.

On the other hand, if excitons and carriers are created with linearly polarised light  or are otherwise created without a significant spin polarisation, then all of the spin scattering events in Table \ref{table:S_exch^Xeh} are possible, and the carriers populate the states $\ket{\pm 1/2} (\ket{\pm 3/2})$ equally. Denoting the total electron density as $n_e$, for we get the following  expression for the broadening of the exciton resonance:
\begin{align} \label{eqn:Gamma_Xe_f_lin}
    \Gamma_{X{\text-}e}  = \sum\limits_{\substack{S_{e'},\\ S_X^f, S_{e'}^f}} \frac{n_e}{2} 
    \int_{0}^{\infty}&{\rm d}q\   q    \left|  J_{S_X S_{e'}}^{S_X^f S_{e'}^{f}} (0, 0, q) \right|^2  \nonumber\\[-.5cm]
    &\times \frac{1}{\pi} \frac{2\Gamma_{X{\text-}e}}{\left(\frac{\hbar^2 q^2}{2M^*_e}\right)^2 + 4\Gamma_{X{\text-}e}^2}.
\end{align}

\begin{figure*}[ht]
    \centering
    \includegraphics[]{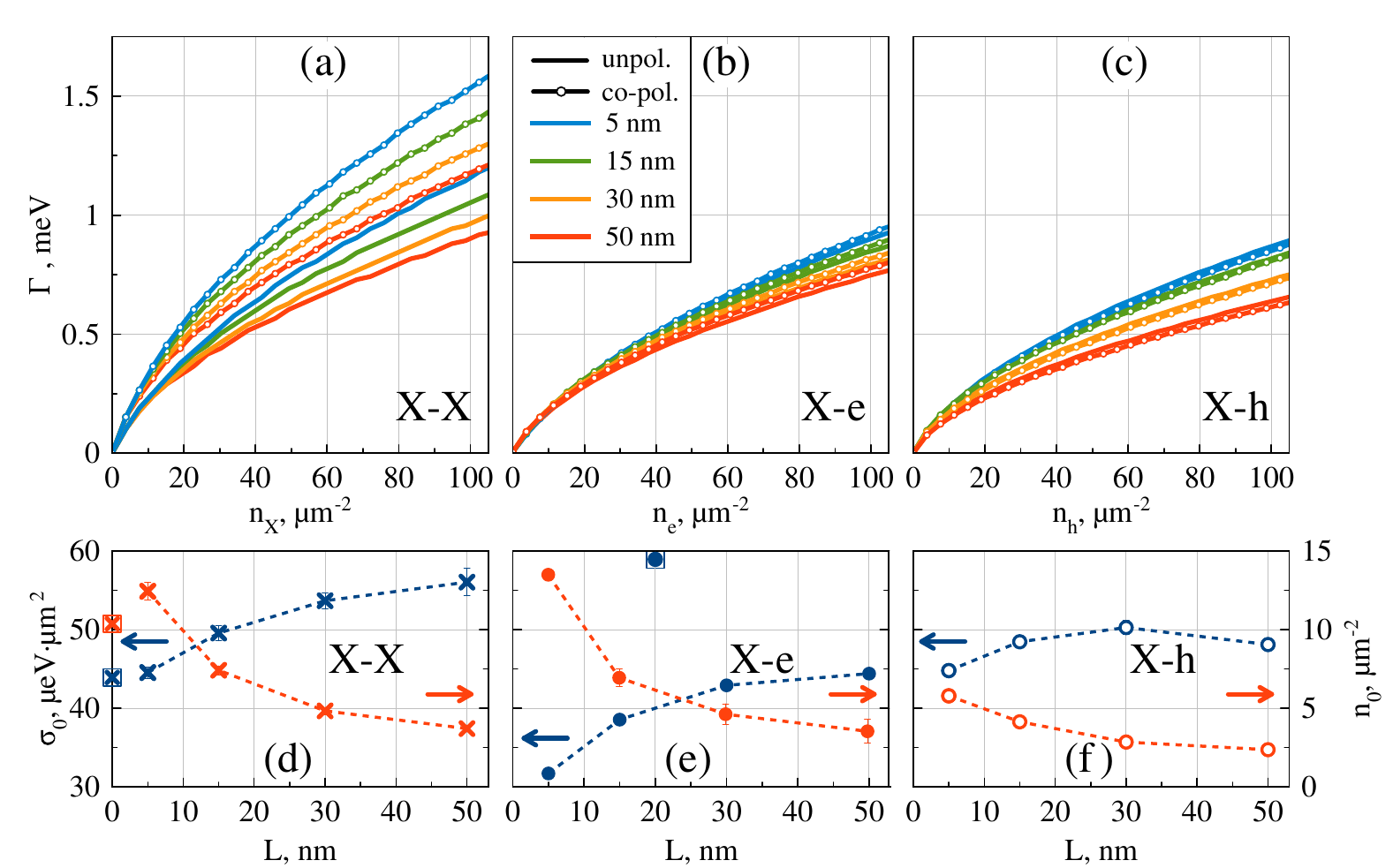}
    \caption{Top panels: density-dependent collisional broadening of the heavy-hole exciton resonance in square GaAs/Al$_{0.3}$Ga$_{0.7}$As QWs for three scattering mechanisms: (a) exciton-exciton scattering, (b) exciton-electron scattering, (c) exciton-hole scattering. Solid lines: all spin states populated equally  (cold exciton reservoir or luminescence experiments), see Eqs.~(\ref{eqn:Gamma_XX_f_lin}), (\ref{eqn:Gamma_Xe_f_lin});  lines with circles: only the spin states $\ket{+1}$  (excitons), $\ket{-1/2}$  (electrons) and $\ket{+3/2}$  (holes), or the equivalent opposite sign states are populated,  and the probe pulse is co-polarised, see Eqs.~(\ref{eqn:Gamma_XX_f_circ}), (\ref{eqn:Gamma_Xe_f_circ}). Note: 100~$\mu$m$^{-2} = 1 \cdot 10^{10}$~cm$^{-2}$. The bottom panels (d), (e), and (f) display the $L$ dependences of the low-density cross-section $\sigma_0$ (blue symbols, left axes) and the critical density $n_0$ (red symbols, right axes) for the three scattering processes  in the unpolarised case  (see Eq.~(\ref{eqn:fit_func})). Squared crosses in panel (d) are results of fitting the unpolarised curves from Ref.~\cite{ciuti1998}, the squared solid circle in panel (e) is the linear cross-section extracted from the 5~K curves of Ref.~\cite{cohen_neutral&charged_exciton-electron2003}.}
    \label{fig:Gamma}
\end{figure*}

We use the expressions (\ref{eqn:Gamma_XX_f_circ})-(\ref{eqn:Gamma_Xe_f_lin}) to calculate the density-dependent collisional broadening for four GaAs/Al$_{0.3}$Ga$_{0.7}$As QWs ($L = $ 5, 15, 30, 50~nm) due to exciton-exciton, exciton-electron and exciton-hole scattering and in both the co-polarised and the unpolarised cases. The results are presented in Fig.~\ref{fig:Gamma}, see panels (a), (b) and (c). 

It is immediately apparent that all interactions grow in strength with decreasing QW width $L$. The reason for this is evident. With decreasing $L$, the exciton is squeezed not only along the growth axis, but also in the $(x, y)$ plane, and in the limiting 2D-case, the exciton Bohr radius is reduced to $a_B/2$. The increased confinement is connected to increased delocalisation of the exciton in reciprocal space, which results in the widening of the q-dependences with decreasing $L$ in Fig.~\ref{fig:J_q_tryptich}. In turn, this increases the integrals in the broadening equations  (\ref{eqn:Gamma_XX_f_circ})-(\ref{eqn:Gamma_Xe_f_lin}), leading to an increase in the broadening. One may treat this effect as a manifestation of the uncertainty principle: increased confinement leads to greater momentum uncertainty, which translates to greater broadening of the energy levels. The exciton-exciton matrix elements are stronger affected by this exciton squeezing than the exciton-carrier ones (see Fig.~\ref{fig:J_q_tryptich}), which leads to a strong dependence on $L$ in Fig.~\ref{fig:Gamma}(a). In the exciton-electron case the $L$-dependence is weakened because the direct and exchange terms partially compensate each other in the region of large $q$ (see Fig.~\ref{fig:J_q_tryptich}(b)).


With $L$ decreasing further, the overall magnitudes of the matrix elements begin to decrease as a consequence of the wave functions penetrating the barriers. The decrease at all values of $q$ is approximately proportional to the decrease of the exchange constants in Fig.~\ref{fig:J_all}. Once this decrease becomes large enough, it compensates the increasing width of the $J(q)$ curves. In the X-e and X-h cases, where the exchange constants experience a sharper decrease compared to the X-X case (see Fig.~\ref{fig:J_all}), the beginning of a saturation-like behaviour of $\Gamma$ is observed in narrow QWs. We have also studied the $\Gamma(n)$ dependences in a 3~nm QW, though the use of an effective mass approach in such narrow QWs is debatable. Nevertheless, in these calculations we have observed a decrease in $\Gamma$ relative to the 5~nm case in all three types of scattering, caused by the significant barrier penetration of the exciton and carrier wave functions.


The exciton-exciton collisional broadening is noticeably different in the co-polarised and unpolarised cases, whereas exciton-carrier scattering is almost independent of polarisation. This is explained by a larger number of spin scattering channels for the unpolarised exciton-exciton scattering. We have also considered various cases of partial spin polarisation of excitons and carriers. As expected, the broadening turned out to be generally greater than in the unpolarised case, but less than in the case of complete polarisation.

All of the broadening curves presented in Fig.~\ref{fig:Gamma}(a), (b), and (c) demonstrate sublinear dependence on the exciton or carrier areal density.  This is an effect caused by the Lorentzian in Eqs.~(\ref{eqn:Gamma_XX_f_lin})-(\ref{eqn:Gamma_Xe_f_circ}): as $\Gamma$ increases, scattering events with greater transferred momentum $q$ begin to occur. The matrix elements near $q = 0$ decrease with rising $q$, and this causes the sublinear dependence. The sublinear behaviour of the curves can be easily analysed. In the limit of very small broadening, the Lorentzian behaves similarly to a $\delta$-function. This eliminates the dependence on $\Gamma$ on the right side of the equations, leading to a linear $\Gamma(n)$ dependence near $n=0$  with a coefficient $\sigma_0$ that is proportional to the corresponding exchange constant squared. In the exciton-exciton case, for example, Eqs.~(\ref{eqn:Gamma_XX_f_circ}) and (\ref{eqn:Gamma_XX_f_lin}) dictate that the relationships between $\sigma_0$ and the exchange constants $J_{exch}$ are $\sigma_0 = 4 J_{exch}^2$ and $\sigma_0 = 2.5 J_{exch}^2$ in the co-polarised and unpolarised cases, respectively. At large $\Gamma$, when the Lorentzian is much broader in $q$ than the matrix elements $J(0,0,q)$, the asymptotic behaviour of the broadening is such that $\Gamma(n) \sim \sqrt{n}$, with the coefficient being proportional to the integral of $J_{\rm exch}^2(0,0,q)$. These considerations allow fitting the curves with the two-parameter functions 
\begin{align} \label{eqn:fit_func}
\Gamma = \sigma_0 \frac{n}{1 + \sqrt{n/n_0}}, 
\end{align}
which accurately model the numerical results, with relative errors of the fitting parameters $\sigma_0$, $n_0$ in the order of $1 \%$. Based on Eq.~(\ref{eqn:fit_func}), we may introduce the density-dependent scattering cross-section
\begin{align} \label{eqn:sigma_n}
    \sigma(n) =  \frac{\sigma_0}{1 + \sqrt{n/n_0}}.
\end{align}
\newcommand*\rfrac[2]{{}^{#1}\!/_{#2}}
The physical meaning of the fitting parameters  is evident from Eq.~(\ref{eqn:sigma_n}): $\sigma_0$ is the scattering cross-section in the linear regime, when $\Gamma(n) = \sigma_0 n$, and $n_0$ is the quasiparticle density at which the cross-section is halved, i.e. $\sigma(n_0) = \sigma_0 / 2$. This parameter determines the density at which the $\Gamma(n)$ dependence becomes essentially sublinear. The fit results in the unpolarised case are plotted in panels (d), (e) and (f) of Fig.~\ref{fig:Gamma}. The low-density cross-sections $\sigma_0$ vary within the range 30--60~$\mu$eV$\cdot\mu$m$^2$. Since they are proportional to the corresponding exchange constants squared, their QW dependences closely follow the dependences in Fig.~\ref{fig:J_all}, with the ratio $2.5\, J_{\rm exch}^2 / \sigma_0 \approx 1.2\,\text{-}\,1.3$. This is a consequence of the true linear regime actually being restricted by infinitesimally small $n$, which itself is a consequence of  ${\rm d}\sigma/{\rm d} n \neq 0$ near $n=0$. Because of this, the discrete nature of the $\Gamma(n)$ dependences leads to an underestimate of the linear cross-section. It may also be partly attributed to the choice of the model function~(\ref{eqn:fit_func}). The same behaviour is observed in our fit data for the co-polarised case, with $4\, J_{\rm exch}^2 / \sigma_0 \approx 1.2\,\text{-}\,1.3$. 

In wide QWs, the critical densities $n_0$ are quite small, with $n_0 < 5$~$\mu$m$^{-2}$ for $L \geq 30$~nm. A significant increase in $n_0$ is observed across all cases in narrow wells, which is a consequence of the excitons being squeezed in the QW plane. This increase in $n_0$ widens the density range where the broadening may be considered linear in density.  We must note that in real experiments the transition from the linear to the sublinear regime may be obscured by noise, as well as errors in determining the exciton/carrier densities. Moreover, the $\Gamma(n)$ dependence may seem linear if the measurements are made across a range of one order of magnitude in density, meaning that special attention must be paid in experiments with seemingly linear power dependences of the exciton resonance broadening.

The exciton-exciton collisional broadening can be compared to the work of C. Ciuti {\it et. al}~\cite{ciuti1998}. We have fitted the unpolarised curves presented in their work with the function~(\ref{eqn:fit_func}), the results are plotted in Fig.~\ref{fig:Gamma}(a). The cross-section in the linear regime is consistent with our results for narrow QWs, as is the critical density. Overall, the exciton-exciton broadening obtained in our work for the narrowest QW ($L = 5$~nm) is larger by a factor of 1.15 than the result of Ref.~\cite{ciuti1998}. There are several possible reasons for this, e.g. the different material parameters used in our work and in theirs, and the difference between our microscopically calculated wave functions and the strictly 2D analytical wave functions of Ref.~\cite{ciuti1998}. Overall, we find our exciton-exciton calculation in good agreement with this work, predictably. 

Our exciton-electron data can be compared to the results obtained for an infinite-barrier 20~nm QW by G. Ramon, A. Mann and E. Cohen~\cite{cohen_neutral&charged_exciton-electron2003}. They take the thermal distribution of electrons into account, while we only consider stationary carriers at very low temperatures. However, they note that the scattering favours events such that $\Delta{\bf Q}_{X{\text -}e} = -{\bf q}$ and that the amplitude only weakly depends on $\Delta{\bf Q}$ in a wide range. Considering this fact, one would expect similar results with our approach. However, our calculations produce a broadening that is smaller than their 5~K result by a factor of 1.5 in the linear regime (see Fig.~\ref{fig:Gamma}(b)), and is much more sublinear at larger electron densities. We ascribe the large difference to the different approaches of our work and Ref.~\cite{cohen_neutral&charged_exciton-electron2003}: we use the self-consistent Eq.~(\ref{eqn:Gamma_Xe_1}), while Fermi's golden rule was used in~\cite{cohen_neutral&charged_exciton-electron2003}, which corresponds to a Lorentzian of zero width in our model. Naturally, this causes our approach to produce a much smaller broadening. 

\section{Conclusions}
\label{sec:conclusions}

In this work, we have developed a general theoretical approach for study of the exciton-exciton and exciton-carrier interaction in quantum wells. We have considered the case of relatively weak interaction, when the unperturbed wave functions of the interacting quasiparticles can be used in the theoretical modelling, and correlation effects can be neglected. The exciton wave functions were calculated by the direct numerical solution of the three-dimensional Schr{\"o}dinger equation for excitons in QWs of different widths with finite potential barriers. The direct Coulomb and exchange matrix elements are obtained by the numerical calculations of the the 9- and 12-dimensional integrals. The spin degrees of freedom are also taken into account via a Hartree-Fock approach. In this work we have considered only the case of small exciton and carrier wave vectors in the QW layer, with a finite transferred momentum $q$ due to the collisions between quasiparticles. Similarly to Refs.~\cite{ciuti1998, cohen_neutral&charged_exciton-electron2003, schindler2008}, we have found that the exchange interaction dominates over the direct one.

We have also calculated the collision-induced broadening of exciton resonances, which is observable in optical experiments. In these calculations we have used an approach based on the second-order Born approximation, taking into account the spin dependence of the interactions.
We have determined that the fermion exchange constants, i.e. the exchange matrix elements at zero momenta, are the greatest factor determining the overall strength of the interactions, particularly at low densities. The other important factor is the confinement of the electron-hole pair inside the exciton. The behaviour of the broadening with respect to QW width is almost entirely determined by the interplay of these two factors. This is supported by analysing the results of fitting the broadening curves in terms of model two-parameter functions. Additionally, the introduction of such functions should help compare our results to experimental works in the future.

We have shown that the exciton-exciton interaction leads to greater broadening than exciton-carrier interaction in the considered model. The QW width dependence is predictably stronger in the exciton-exciton case, and is weakest in the exciton-electron case because of the direct and exchange matrix elements partially compensating each other. 
The different interactions also behave differently with respect to spin polarisation. Exciton-exciton collisional broadening is enhanced by approximately 30\% in the co-polarised case relative to the unpolarised one at large exciton densities  i.e. when $n\in(40; 100)$~$\mu$m$^{-2}$. At the same time, exciton-carrier interactions lead to a broadening that is nearly independent on polarisation. We interpret this as a consequence of the richer spin scattering channels in the exciton-exciton case.

 We have also shown that the cross-section of the exciton-exciton collisional broadening in the low-density regime is proportional to the exchange constant squared, with a coefficient ranging from $2.5\,$-$\,4$, depending on the degree of spin polarisation of the exciton ensemble. We must stress, however, that this relation holds only at very low exciton densities. The sublinear dependence of the broadening on particle density becomes apparent only across a very wide range of densities, around two orders of magnitude. Because of various factors, the sublinearity may be masked in experiments if measurements are made across a small range of densities, making the dependence seem linear. In that case a linear approximation could yield only a very rough estimate of the exchange constants.

We have compared our results to other theoretical works~\cite{ciuti1998, cohen_neutral&charged_exciton-electron2003}. The agreement is deemed adequate, considering the approximations used. Additional studies are required to test the proposed dependences of the broadening on QW width.

Our results indicate that the lineshift (signified by the exchange constants) and the exciton line broadening reach their maxima in QWs of different width. The high-density broadening is maximised in 5-10~nm-QWs (see Fig.~\ref{fig:Gamma}), while the lineshift is greatest in wide 35-75~nm-QWs. This could help choose the best QW structure for a given application.

\section{Acknowledgements}
\label{sec:acknowledgements}
The authors acknowledge K.V.~Kavokin and D.S.~Smirnov for fruitful discussions. This work was supported by the Russian Science Foundation, Grant No. 19-72-20039, which is highly appreciated.


\appendix
\section{Changing the variables}
\label{sec:appendixA}
For the exciton-electron integrals:
\begin{align}
({\bf r}_e, {\bf r}_h, {\bf r}_{e'}) \rightarrow ({\pmb \rho}_{eh}, z_e, z_h, {\pmb \rho}_{e'h}, z_{e'}, {\pmb \sigma}),
\end{align}
where ${\pmb \rho}_{eh(e'h)} = {\bf r}_{e(e')}^\perp - {\bf r}_{h}^\perp$ are the in-plane coordinates of the electrons relative to the hole, ${\pmb \sigma} = (m_e{\bf r}_e^\perp + m_e{\bf r}_{e'}^\perp + m_h{\bf r}_h^\perp) / (2m_e + m_h)$ is the system's centre of mass. Due to the translational symmetry of the exciton-electron system, the integrals do not depend on $\pmb \sigma$, so the integration over $\pmb \sigma$ is trivial and gives the normalization area $A$. This reduces the integrals to 7-dimensional ones. The rest of the coordinates are chosen so that the wave functions vanish as any of them go to infinity. We further increase the convergence of the integral by switching to double polar coordinates in $({\pmb \rho}_{eh}, {\pmb \rho}_{e'h})$. This is due to the fact that uniformly distributed points on a polar grid have a higher density near the origin; in our case, the integrands are largest at the origin and vanish exponentially, which provides enhanced convergence compared to cartesian coordinates.

To calculate the exciton-exciton matrix elements, the variables are also changed:
\begin{align}
({\bf r}_e, {\bf r}_h, {\bf r}_{e'}, {\bf r}_{h'}) \rightarrow ({\pmb \rho}_{eh}, z_e, z_h, {\pmb \rho}_{e'h'}, z_{e'}, z_{h'}, {\pmb \xi}, {\pmb \sigma}),
\end{align}
where ${\pmb \rho}_{eh(e'h')} = {\bf r}_{e(e')}^\perp - {\bf r}_{h(h')}^\perp$ are the in-plane relative coordinates of the excitons, ${\pmb \sigma} = ({\bf R}_{eh} + {\bf R}_{e'h'}) / 2$ is the system's centre of mass, and ${\pmb \xi} = {\bf R}_{eh} - {\bf R}_{e'h'}$ is the distance between the two excitons. Due to the translational symmetry of the two-exciton system, the integrals do not depend on $\pmb \sigma$, so the integration over $\pmb \sigma$ is trivial and gives the normalization area $A$. This reduces the integrals to 10-dimensional ones. Again, the convergence of the integral is further increased by switching to triple polar coordinates in $({\pmb \rho}_{eh}, {\pmb \rho}_{e'h'}, {\pmb \xi})$.


\begin{thebibliography}{59}%
\makeatletter
\providecommand \@ifxundefined [1]{%
 \@ifx{#1\undefined}
}%
\providecommand \@ifnum [1]{%
 \ifnum #1\expandafter \@firstoftwo
 \else \expandafter \@secondoftwo
 \fi
}%
\providecommand \@ifx [1]{%
 \ifx #1\expandafter \@firstoftwo
 \else \expandafter \@secondoftwo
 \fi
}%
\providecommand \natexlab [1]{#1}%
\providecommand \enquote  [1]{``#1''}%
\providecommand \bibnamefont  [1]{#1}%
\providecommand \bibfnamefont [1]{#1}%
\providecommand \citenamefont [1]{#1}%
\providecommand \href@noop [0]{\@secondoftwo}%
\providecommand \href [0]{\begingroup \@sanitize@url \@href}%
\providecommand \@href[1]{\@@startlink{#1}\@@href}%
\providecommand \@@href[1]{\endgroup#1\@@endlink}%
\providecommand \@sanitize@url [0]{\catcode `\\12\catcode `\$12\catcode
  `\&12\catcode `\#12\catcode `\^12\catcode `\_12\catcode `\%12\relax}%
\providecommand \@@startlink[1]{}%
\providecommand \@@endlink[0]{}%
\providecommand \url  [0]{\begingroup\@sanitize@url \@url }%
\providecommand \@url [1]{\endgroup\@href {#1}{\urlprefix }}%
\providecommand \urlprefix  [0]{URL }%
\providecommand \Eprint [0]{\href }%
\providecommand \doibase [0]{https://doi.org/}%
\providecommand \selectlanguage [0]{\@gobble}%
\providecommand \bibinfo  [0]{\@secondoftwo}%
\providecommand \bibfield  [0]{\@secondoftwo}%
\providecommand \translation [1]{[#1]}%
\providecommand \BibitemOpen [0]{}%
\providecommand \bibitemStop [0]{}%
\providecommand \bibitemNoStop [0]{.\EOS\space}%
\providecommand \EOS [0]{\spacefactor3000\relax}%
\providecommand \BibitemShut  [1]{\csname bibitem#1\endcsname}%
\let\auto@bib@innerbib\@empty
\bibitem [{\citenamefont {Ivchenko}(2005)}]{ivchenko2005book}%
  \BibitemOpen
  \bibfield  {author} {\bibinfo {author} {\bibfnamefont {E.~L.}\ \bibnamefont
  {Ivchenko}},\ }\href@noop {} {\emph {\bibinfo {title} {Optical spectroscopy
  of semiconductor nanostructures}}}\ (\bibinfo  {publisher} {Alpha Science
  International Ltd.},\ \bibinfo {year} {2005})\BibitemShut {NoStop}%
\bibitem [{\citenamefont {Khramtsov}\ \emph {et~al.}(2016)\citenamefont
  {Khramtsov}, \citenamefont {Belov}, \citenamefont {Grigoryev}, \citenamefont
  {Ignatiev}, \citenamefont {Verbin}, \citenamefont {Efimov}, \citenamefont
  {Eliseev}, \citenamefont {Lovtcius}, \citenamefont {Petrov},\ and\
  \citenamefont {Yakovlev}}]{khramtsov2016}%
  \BibitemOpen
  \bibfield  {author} {\bibinfo {author} {\bibfnamefont {E.~S.}\ \bibnamefont
  {Khramtsov}}, \bibinfo {author} {\bibfnamefont {P.~A.}\ \bibnamefont
  {Belov}}, \bibinfo {author} {\bibfnamefont {P.~S.}\ \bibnamefont
  {Grigoryev}}, \bibinfo {author} {\bibfnamefont {I.~V.}\ \bibnamefont
  {Ignatiev}}, \bibinfo {author} {\bibfnamefont {S.~Y.}\ \bibnamefont
  {Verbin}}, \bibinfo {author} {\bibfnamefont {Y.~P.}\ \bibnamefont {Efimov}},
  \bibinfo {author} {\bibfnamefont {S.~A.}\ \bibnamefont {Eliseev}}, \bibinfo
  {author} {\bibfnamefont {V.~A.}\ \bibnamefont {Lovtcius}}, \bibinfo {author}
  {\bibfnamefont {V.~V.}\ \bibnamefont {Petrov}},\ and\ \bibinfo {author}
  {\bibfnamefont {S.~L.}\ \bibnamefont {Yakovlev}},\ }\bibfield  {title}
  {\bibinfo {title} {Radiative decay rate of excitons in square quantum wells:
  Microscopic modeling and experiment},\ }\href
  {https://doi.org/10.1063/1.4948664} {\bibfield  {journal} {\bibinfo
  {journal} {Journal of Applied Physics}\ }\textbf {\bibinfo {volume} {119}},\
  \bibinfo {pages} {184301} (\bibinfo {year} {2016})}\BibitemShut {NoStop}%
\bibitem [{\citenamefont {Andreani}\ \emph {et~al.}(1991)\citenamefont
  {Andreani}, \citenamefont {Tassone},\ and\ \citenamefont
  {Bassani}}]{andreani1991-radiative-lifetime}%
  \BibitemOpen
  \bibfield  {author} {\bibinfo {author} {\bibfnamefont {L.~C.}\ \bibnamefont
  {Andreani}}, \bibinfo {author} {\bibfnamefont {F.}~\bibnamefont {Tassone}},\
  and\ \bibinfo {author} {\bibfnamefont {F.}~\bibnamefont {Bassani}},\
  }\bibfield  {title} {\bibinfo {title} {Radiative lifetime of free excitons in
  quantum wells},\ }\href
  {https://doi.org/https://doi.org/10.1016/0038-1098(91)90761-J} {\bibfield
  {journal} {\bibinfo  {journal} {Solid State Communications}\ }\textbf
  {\bibinfo {volume} {77}},\ \bibinfo {pages} {641} (\bibinfo {year}
  {1991})}\BibitemShut {NoStop}%
\bibitem [{\citenamefont {D’Andrea}\ \emph {et~al.}(1998)\citenamefont
  {D’Andrea}, \citenamefont {Tomassini}, \citenamefont {Ferrari},
  \citenamefont {Righini}, \citenamefont {Selci}, \citenamefont {Bruni},
  \citenamefont {Schiumarini},\ and\ \citenamefont
  {Simeone}}]{d'andrea1998-radiative-lifetime}%
  \BibitemOpen
  \bibfield  {author} {\bibinfo {author} {\bibfnamefont {A.}~\bibnamefont
  {D’Andrea}}, \bibinfo {author} {\bibfnamefont {N.}~\bibnamefont
  {Tomassini}}, \bibinfo {author} {\bibfnamefont {L.}~\bibnamefont {Ferrari}},
  \bibinfo {author} {\bibfnamefont {M.}~\bibnamefont {Righini}}, \bibinfo
  {author} {\bibfnamefont {S.}~\bibnamefont {Selci}}, \bibinfo {author}
  {\bibfnamefont {M.~R.}\ \bibnamefont {Bruni}}, \bibinfo {author}
  {\bibfnamefont {D.}~\bibnamefont {Schiumarini}},\ and\ \bibinfo {author}
  {\bibfnamefont {M.~G.}\ \bibnamefont {Simeone}},\ }\bibfield  {title}
  {\bibinfo {title} {Optical properties of stepped
  {I}n$_x${G}a$_{1−x}${A}s/{G}a{A}s quantum wells},\ }\href
  {https://doi.org/10.1063/1.367971} {\bibfield  {journal} {\bibinfo  {journal}
  {Journal of Applied Physics}\ }\textbf {\bibinfo {volume} {83}},\ \bibinfo
  {pages} {7920} (\bibinfo {year} {1998})}\BibitemShut {NoStop}%
\bibitem [{\citenamefont {Grigoryev}\ \emph {et~al.}(2016)\citenamefont
  {Grigoryev}, \citenamefont {Kurdyubov}, \citenamefont {Kuznetsova},
  \citenamefont {Ignatiev}, \citenamefont {Efimov}, \citenamefont {Eliseev},
  \citenamefont {Petrov}, \citenamefont {Lovtcius},\ and\ \citenamefont
  {Shapochkin}}]{grigoryev2016}%
  \BibitemOpen
  \bibfield  {author} {\bibinfo {author} {\bibfnamefont {P.}~\bibnamefont
  {Grigoryev}}, \bibinfo {author} {\bibfnamefont {A.}~\bibnamefont
  {Kurdyubov}}, \bibinfo {author} {\bibfnamefont {M.}~\bibnamefont
  {Kuznetsova}}, \bibinfo {author} {\bibfnamefont {I.}~\bibnamefont
  {Ignatiev}}, \bibinfo {author} {\bibfnamefont {Y.}~\bibnamefont {Efimov}},
  \bibinfo {author} {\bibfnamefont {S.}~\bibnamefont {Eliseev}}, \bibinfo
  {author} {\bibfnamefont {V.}~\bibnamefont {Petrov}}, \bibinfo {author}
  {\bibfnamefont {V.}~\bibnamefont {Lovtcius}},\ and\ \bibinfo {author}
  {\bibfnamefont {P.}~\bibnamefont {Shapochkin}},\ }\bibfield  {title}
  {\bibinfo {title} {Excitons in asymmetric quantum wells},\ }\href
  {https://doi.org/https://doi.org/10.1016/j.spmi.2016.07.008} {\bibfield
  {journal} {\bibinfo  {journal} {Superlattices and Microstructures}\ }\textbf
  {\bibinfo {volume} {97}},\ \bibinfo {pages} {452} (\bibinfo {year}
  {2016})}\BibitemShut {NoStop}%
\bibitem [{\citenamefont {Khramtsov}\ \emph {et~al.}(2019)\citenamefont
  {Khramtsov}, \citenamefont {Grigoryev}, \citenamefont {Loginov},
  \citenamefont {Ignatiev}, \citenamefont {Efimov}, \citenamefont {Eliseev},
  \citenamefont {Shapochkin}, \citenamefont {Ivchenko},\ and\ \citenamefont
  {Bayer}}]{khramtsov2019}%
  \BibitemOpen
  \bibfield  {author} {\bibinfo {author} {\bibfnamefont {E.~S.}\ \bibnamefont
  {Khramtsov}}, \bibinfo {author} {\bibfnamefont {P.~S.}\ \bibnamefont
  {Grigoryev}}, \bibinfo {author} {\bibfnamefont {D.~K.}\ \bibnamefont
  {Loginov}}, \bibinfo {author} {\bibfnamefont {I.~V.}\ \bibnamefont
  {Ignatiev}}, \bibinfo {author} {\bibfnamefont {Y.~P.}\ \bibnamefont
  {Efimov}}, \bibinfo {author} {\bibfnamefont {S.~A.}\ \bibnamefont {Eliseev}},
  \bibinfo {author} {\bibfnamefont {P.~Y.}\ \bibnamefont {Shapochkin}},
  \bibinfo {author} {\bibfnamefont {E.~L.}\ \bibnamefont {Ivchenko}},\ and\
  \bibinfo {author} {\bibfnamefont {M.}~\bibnamefont {Bayer}},\ }\bibfield
  {title} {\bibinfo {title} {Exciton spectroscopy of optical reflection from
  wide quantum wells},\ }\href {https://doi.org/10.1103/PhysRevB.99.035431}
  {\bibfield  {journal} {\bibinfo  {journal} {Phys. Rev. B}\ }\textbf {\bibinfo
  {volume} {99}},\ \bibinfo {pages} {035431} (\bibinfo {year}
  {2019})}\BibitemShut {NoStop}%
\bibitem [{\citenamefont {Butov}\ \emph {et~al.}(2002)\citenamefont {Butov},
  \citenamefont {Gossard},\ and\ \citenamefont {Chemla}}]{butov2002}%
  \BibitemOpen
  \bibfield  {author} {\bibinfo {author} {\bibfnamefont {L.~V.}\ \bibnamefont
  {Butov}}, \bibinfo {author} {\bibfnamefont {A.~C.}\ \bibnamefont {Gossard}},\
  and\ \bibinfo {author} {\bibfnamefont {D.~S.}\ \bibnamefont {Chemla}},\
  }\bibfield  {title} {\bibinfo {title} {Macroscopically ordered state in an
  exciton system},\ }\href {https://doi.org/10.1038/nature00943} {\bibfield
  {journal} {\bibinfo  {journal} {Nature}\ }\textbf {\bibinfo {volume} {418}},\
  \bibinfo {pages} {751} (\bibinfo {year} {2002})}\BibitemShut {NoStop}%
\bibitem [{\citenamefont {V\"or\"os}\ \emph {et~al.}(2009)\citenamefont
  {V\"or\"os}, \citenamefont {Snoke}, \citenamefont {Pfeiffer},\ and\
  \citenamefont {West}}]{voros2009}%
  \BibitemOpen
  \bibfield  {author} {\bibinfo {author} {\bibfnamefont {Z.}~\bibnamefont
  {V\"or\"os}}, \bibinfo {author} {\bibfnamefont {D.~W.}\ \bibnamefont
  {Snoke}}, \bibinfo {author} {\bibfnamefont {L.}~\bibnamefont {Pfeiffer}},\
  and\ \bibinfo {author} {\bibfnamefont {K.}~\bibnamefont {West}},\ }\bibfield
  {title} {\bibinfo {title} {Direct measurement of exciton-exciton interaction
  energy},\ }\href {https://doi.org/10.1103/PhysRevLett.103.016403} {\bibfield
  {journal} {\bibinfo  {journal} {Phys. Rev. Lett.}\ }\textbf {\bibinfo
  {volume} {103}},\ \bibinfo {pages} {016403} (\bibinfo {year}
  {2009})}\BibitemShut {NoStop}%
\bibitem [{\citenamefont {Andreakou}\ \emph {et~al.}(2015)\citenamefont
  {Andreakou}, \citenamefont {Cronenberger}, \citenamefont {Scalbert},
  \citenamefont {Nalitov}, \citenamefont {Gippius}, \citenamefont {Kavokin},
  \citenamefont {Nawrocki}, \citenamefont {Leonard}, \citenamefont {Butov},
  \citenamefont {Campman}, \citenamefont {Gossard},\ and\ \citenamefont
  {Vladimirova}}]{butov2015}%
  \BibitemOpen
  \bibfield  {author} {\bibinfo {author} {\bibfnamefont {P.}~\bibnamefont
  {Andreakou}}, \bibinfo {author} {\bibfnamefont {S.}~\bibnamefont
  {Cronenberger}}, \bibinfo {author} {\bibfnamefont {D.}~\bibnamefont
  {Scalbert}}, \bibinfo {author} {\bibfnamefont {A.}~\bibnamefont {Nalitov}},
  \bibinfo {author} {\bibfnamefont {N.~A.}\ \bibnamefont {Gippius}}, \bibinfo
  {author} {\bibfnamefont {A.~V.}\ \bibnamefont {Kavokin}}, \bibinfo {author}
  {\bibfnamefont {M.}~\bibnamefont {Nawrocki}}, \bibinfo {author}
  {\bibfnamefont {J.~R.}\ \bibnamefont {Leonard}}, \bibinfo {author}
  {\bibfnamefont {L.~V.}\ \bibnamefont {Butov}}, \bibinfo {author}
  {\bibfnamefont {K.~L.}\ \bibnamefont {Campman}}, \bibinfo {author}
  {\bibfnamefont {A.~C.}\ \bibnamefont {Gossard}},\ and\ \bibinfo {author}
  {\bibfnamefont {M.}~\bibnamefont {Vladimirova}},\ }\bibfield  {title}
  {\bibinfo {title} {Nonlinear optical spectroscopy of indirect excitons in
  coupled quantum wells},\ }\href {https://doi.org/10.1103/PhysRevB.91.125437}
  {\bibfield  {journal} {\bibinfo  {journal} {Phys. Rev. B}\ }\textbf {\bibinfo
  {volume} {91}},\ \bibinfo {pages} {125437} (\bibinfo {year}
  {2015})}\BibitemShut {NoStop}%
\bibitem [{\citenamefont {Choksy}\ \emph {et~al.}(2021)\citenamefont {Choksy},
  \citenamefont {Xu}, \citenamefont {Fogler}, \citenamefont {Butov},
  \citenamefont {Norman},\ and\ \citenamefont {Gossard}}]{butov2021}%
  \BibitemOpen
  \bibfield  {author} {\bibinfo {author} {\bibfnamefont {D.~J.}\ \bibnamefont
  {Choksy}}, \bibinfo {author} {\bibfnamefont {C.}~\bibnamefont {Xu}}, \bibinfo
  {author} {\bibfnamefont {M.~M.}\ \bibnamefont {Fogler}}, \bibinfo {author}
  {\bibfnamefont {L.~V.}\ \bibnamefont {Butov}}, \bibinfo {author}
  {\bibfnamefont {J.}~\bibnamefont {Norman}},\ and\ \bibinfo {author}
  {\bibfnamefont {A.~C.}\ \bibnamefont {Gossard}},\ }\bibfield  {title}
  {\bibinfo {title} {Attractive and repulsive dipolar interaction in bilayers
  of indirect excitons},\ }\href {https://doi.org/10.1103/PhysRevB.103.045126}
  {\bibfield  {journal} {\bibinfo  {journal} {Phys. Rev. B}\ }\textbf {\bibinfo
  {volume} {103}},\ \bibinfo {pages} {045126} (\bibinfo {year}
  {2021})}\BibitemShut {NoStop}%
\bibitem [{\citenamefont {Kavokin}\ \emph {et~al.}(2017)\citenamefont
  {Kavokin}, \citenamefont {Baumberg}, \citenamefont {Malpuech},\ and\
  \citenamefont {Laussy}}]{kavokin2017microcavities}%
  \BibitemOpen
  \bibfield  {author} {\bibinfo {author} {\bibfnamefont {A.}~\bibnamefont
  {Kavokin}}, \bibinfo {author} {\bibfnamefont {J.~J.}\ \bibnamefont
  {Baumberg}}, \bibinfo {author} {\bibfnamefont {G.}~\bibnamefont {Malpuech}},\
  and\ \bibinfo {author} {\bibfnamefont {F.~P.}\ \bibnamefont {Laussy}},\
  }\href@noop {} {\emph {\bibinfo {title} {Microcavities}}}\ (\bibinfo
  {publisher} {Oxford university press},\ \bibinfo {year} {2017})\BibitemShut
  {NoStop}%
\bibitem [{\citenamefont {Shahnazaryan}\ \emph {et~al.}(2017)\citenamefont
  {Shahnazaryan}, \citenamefont {Iorsh}, \citenamefont {Shelykh},\ and\
  \citenamefont {Kyriienko}}]{shelykh2017}%
  \BibitemOpen
  \bibfield  {author} {\bibinfo {author} {\bibfnamefont {V.}~\bibnamefont
  {Shahnazaryan}}, \bibinfo {author} {\bibfnamefont {I.}~\bibnamefont {Iorsh}},
  \bibinfo {author} {\bibfnamefont {I.~A.}\ \bibnamefont {Shelykh}},\ and\
  \bibinfo {author} {\bibfnamefont {O.}~\bibnamefont {Kyriienko}},\ }\bibfield
  {title} {\bibinfo {title} {Exciton-exciton interaction in transition-metal
  dichalcogenide monolayers},\ }\href
  {https://doi.org/10.1103/PhysRevB.96.115409} {\bibfield  {journal} {\bibinfo
  {journal} {Phys. Rev. B}\ }\textbf {\bibinfo {volume} {96}},\ \bibinfo
  {pages} {115409} (\bibinfo {year} {2017})}\BibitemShut {NoStop}%
\bibitem [{\citenamefont {Erkensten}\ \emph {et~al.}(2021)\citenamefont
  {Erkensten}, \citenamefont {Brem},\ and\ \citenamefont
  {Malic}}]{erkensten2021}%
  \BibitemOpen
  \bibfield  {author} {\bibinfo {author} {\bibfnamefont {D.}~\bibnamefont
  {Erkensten}}, \bibinfo {author} {\bibfnamefont {S.}~\bibnamefont {Brem}},\
  and\ \bibinfo {author} {\bibfnamefont {E.}~\bibnamefont {Malic}},\ }\bibfield
   {title} {\bibinfo {title} {Exciton-exciton interaction in transition metal
  dichalcogenide monolayers and van der waals heterostructures},\ }\href
  {https://doi.org/10.1103/PhysRevB.103.045426} {\bibfield  {journal} {\bibinfo
   {journal} {Phys. Rev. B}\ }\textbf {\bibinfo {volume} {103}},\ \bibinfo
  {pages} {045426} (\bibinfo {year} {2021})}\BibitemShut {NoStop}%
\bibitem [{\citenamefont {Efimkin}\ \emph {et~al.}(2021)\citenamefont
  {Efimkin}, \citenamefont {Laird}, \citenamefont {Levinsen}, \citenamefont
  {Parish},\ and\ \citenamefont {MacDonald}}]{efimkin2021}%
  \BibitemOpen
  \bibfield  {author} {\bibinfo {author} {\bibfnamefont {D.~K.}\ \bibnamefont
  {Efimkin}}, \bibinfo {author} {\bibfnamefont {E.~K.}\ \bibnamefont {Laird}},
  \bibinfo {author} {\bibfnamefont {J.}~\bibnamefont {Levinsen}}, \bibinfo
  {author} {\bibfnamefont {M.~M.}\ \bibnamefont {Parish}},\ and\ \bibinfo
  {author} {\bibfnamefont {A.~H.}\ \bibnamefont {MacDonald}},\ }\bibfield
  {title} {\bibinfo {title} {Electron-exciton interactions in the
  exciton-polaron problem},\ }\href
  {https://doi.org/10.1103/PhysRevB.103.075417} {\bibfield  {journal} {\bibinfo
   {journal} {Phys. Rev. B}\ }\textbf {\bibinfo {volume} {103}},\ \bibinfo
  {pages} {075417} (\bibinfo {year} {2021})}\BibitemShut {NoStop}%
\bibitem [{\citenamefont {Magde}\ and\ \citenamefont {Mahr}(1970)}]{other1}%
  \BibitemOpen
  \bibfield  {author} {\bibinfo {author} {\bibfnamefont {D.}~\bibnamefont
  {Magde}}\ and\ \bibinfo {author} {\bibfnamefont {H.}~\bibnamefont {Mahr}},\
  }\bibfield  {title} {\bibinfo {title} {Exciton-exciton interaction in cds,
  cdse, and zno},\ }\href {https://doi.org/10.1103/PhysRevLett.24.890}
  {\bibfield  {journal} {\bibinfo  {journal} {Phys. Rev. Lett.}\ }\textbf
  {\bibinfo {volume} {24}},\ \bibinfo {pages} {890} (\bibinfo {year}
  {1970})}\BibitemShut {NoStop}%
\bibitem [{\citenamefont {Piryatinski}\ \emph {et~al.}(2007)\citenamefont
  {Piryatinski}, \citenamefont {Ivanov}, \citenamefont {Tretiak},\ and\
  \citenamefont {Klimov}}]{other2}%
  \BibitemOpen
  \bibfield  {author} {\bibinfo {author} {\bibfnamefont {A.}~\bibnamefont
  {Piryatinski}}, \bibinfo {author} {\bibfnamefont {S.~A.}\ \bibnamefont
  {Ivanov}}, \bibinfo {author} {\bibfnamefont {S.}~\bibnamefont {Tretiak}},\
  and\ \bibinfo {author} {\bibfnamefont {V.~I.}\ \bibnamefont {Klimov}},\
  }\bibfield  {title} {\bibinfo {title} {Effect of quantum and dielectric
  confinement on the exciton−exciton interaction energy in type ii core/shell
  semiconductor nanocrystals},\ }\href {https://doi.org/10.1021/nl0622404}
  {\bibfield  {journal} {\bibinfo  {journal} {Nano Letters}\ }\textbf {\bibinfo
  {volume} {7}},\ \bibinfo {pages} {108} (\bibinfo {year} {2007})}\BibitemShut
  {NoStop}%
\bibitem [{\citenamefont {Kirm}\ \emph {et~al.}(2009)\citenamefont {Kirm},
  \citenamefont {Nagirnyi}, \citenamefont {Feldbach}, \citenamefont
  {De~Grazia}, \citenamefont {Carr\'e}, \citenamefont {Merdji}, \citenamefont
  {Guizard}, \citenamefont {Geoffroy}, \citenamefont {Gaudin}, \citenamefont
  {Fedorov}, \citenamefont {Martin}, \citenamefont {Vasil'ev},\ and\
  \citenamefont {Belsky}}]{other3}%
  \BibitemOpen
  \bibfield  {author} {\bibinfo {author} {\bibfnamefont {M.}~\bibnamefont
  {Kirm}}, \bibinfo {author} {\bibfnamefont {V.}~\bibnamefont {Nagirnyi}},
  \bibinfo {author} {\bibfnamefont {E.}~\bibnamefont {Feldbach}}, \bibinfo
  {author} {\bibfnamefont {M.}~\bibnamefont {De~Grazia}}, \bibinfo {author}
  {\bibfnamefont {B.}~\bibnamefont {Carr\'e}}, \bibinfo {author} {\bibfnamefont
  {H.}~\bibnamefont {Merdji}}, \bibinfo {author} {\bibfnamefont
  {S.}~\bibnamefont {Guizard}}, \bibinfo {author} {\bibfnamefont
  {G.}~\bibnamefont {Geoffroy}}, \bibinfo {author} {\bibfnamefont
  {J.}~\bibnamefont {Gaudin}}, \bibinfo {author} {\bibfnamefont
  {N.}~\bibnamefont {Fedorov}}, \bibinfo {author} {\bibfnamefont
  {P.}~\bibnamefont {Martin}}, \bibinfo {author} {\bibfnamefont
  {A.}~\bibnamefont {Vasil'ev}},\ and\ \bibinfo {author} {\bibfnamefont
  {A.}~\bibnamefont {Belsky}},\ }\bibfield  {title} {\bibinfo {title}
  {Exciton-exciton interactions in ${\text{cdwo}}_{4}$ irradiated by intense
  femtosecond vacuum ultraviolet pulses},\ }\href
  {https://doi.org/10.1103/PhysRevB.79.233103} {\bibfield  {journal} {\bibinfo
  {journal} {Phys. Rev. B}\ }\textbf {\bibinfo {volume} {79}},\ \bibinfo
  {pages} {233103} (\bibinfo {year} {2009})}\BibitemShut {NoStop}%
\bibitem [{\citenamefont {Spector}\ \emph {et~al.}(1986)\citenamefont
  {Spector}, \citenamefont {Lee},\ and\ \citenamefont
  {Melman}}]{spectorLeeMelman1986e-ph}%
  \BibitemOpen
  \bibfield  {author} {\bibinfo {author} {\bibfnamefont {H.~N.}\ \bibnamefont
  {Spector}}, \bibinfo {author} {\bibfnamefont {J.}~\bibnamefont {Lee}},\ and\
  \bibinfo {author} {\bibfnamefont {P.}~\bibnamefont {Melman}},\ }\bibfield
  {title} {\bibinfo {title} {Exciton linewidth in semiconducting quantum-well
  structures},\ }\href {https://doi.org/10.1103/PhysRevB.34.2554} {\bibfield
  {journal} {\bibinfo  {journal} {Phys. Rev. B}\ }\textbf {\bibinfo {volume}
  {34}},\ \bibinfo {pages} {2554} (\bibinfo {year} {1986})}\BibitemShut
  {NoStop}%
\bibitem [{\citenamefont {Grigorchuk}(1999)}]{grigorchuk1999e-ph}%
  \BibitemOpen
  \bibfield  {author} {\bibinfo {author} {\bibfnamefont {N.~I.}\ \bibnamefont
  {Grigorchuk}},\ }\bibfield  {title} {\bibinfo {title} {Exciton-phonon
  coupling and exciton damping due to acoustic phonons in anisotropic nonpolar
  crystals},\ }\href {https://doi.org/10.1088/0953-8984/11/2/008} {\bibfield
  {journal} {\bibinfo  {journal} {Journal of Physics: Condensed Matter}\
  }\textbf {\bibinfo {volume} {11}},\ \bibinfo {pages} {417} (\bibinfo {year}
  {1999})}\BibitemShut {NoStop}%
\bibitem [{\citenamefont {Gopal}\ \emph {et~al.}(2000)\citenamefont {Gopal},
  \citenamefont {Kumar}, \citenamefont {Vengurlekar}, \citenamefont {Bosacchi},
  \citenamefont {Franchi},\ and\ \citenamefont {Pfeiffer}}]{gopal2000e-ph}%
  \BibitemOpen
  \bibfield  {author} {\bibinfo {author} {\bibfnamefont {A.~V.}\ \bibnamefont
  {Gopal}}, \bibinfo {author} {\bibfnamefont {R.}~\bibnamefont {Kumar}},
  \bibinfo {author} {\bibfnamefont {A.~S.}\ \bibnamefont {Vengurlekar}},
  \bibinfo {author} {\bibfnamefont {A.}~\bibnamefont {Bosacchi}}, \bibinfo
  {author} {\bibfnamefont {S.}~\bibnamefont {Franchi}},\ and\ \bibinfo {author}
  {\bibfnamefont {L.~N.}\ \bibnamefont {Pfeiffer}},\ }\bibfield  {title}
  {\bibinfo {title} {Photoluminescence study of exciton–optical phonon
  scattering in bulk {G}a{A}s and {G}a{A}s quantum wells},\ }\href
  {https://doi.org/10.1063/1.372104} {\bibfield  {journal} {\bibinfo  {journal}
  {Journal of Applied Physics}\ }\textbf {\bibinfo {volume} {87}},\ \bibinfo
  {pages} {1858} (\bibinfo {year} {2000})}\BibitemShut {NoStop}%
\bibitem [{\citenamefont {Zhao}\ \emph {et~al.}(2002)\citenamefont {Zhao},
  \citenamefont {Wachter},\ and\ \citenamefont {Kalt}}]{zhao2002e-ph}%
  \BibitemOpen
  \bibfield  {author} {\bibinfo {author} {\bibfnamefont {H.}~\bibnamefont
  {Zhao}}, \bibinfo {author} {\bibfnamefont {S.}~\bibnamefont {Wachter}},\ and\
  \bibinfo {author} {\bibfnamefont {H.}~\bibnamefont {Kalt}},\ }\bibfield
  {title} {\bibinfo {title} {Effect of quantum confinement on exciton-phonon
  interactions},\ }\href {https://doi.org/10.1103/PhysRevB.66.085337}
  {\bibfield  {journal} {\bibinfo  {journal} {Phys. Rev. B}\ }\textbf {\bibinfo
  {volume} {66}},\ \bibinfo {pages} {085337} (\bibinfo {year}
  {2002})}\BibitemShut {NoStop}%
\bibitem [{\citenamefont {Thr\"anhardt}\ \emph {et~al.}(2003)\citenamefont
  {Thr\"anhardt}, \citenamefont {Ell}, \citenamefont {Mosor}, \citenamefont
  {Rupper}, \citenamefont {Khitrova}, \citenamefont {Gibbs},\ and\
  \citenamefont {Koch}}]{thranhardt2003e-ph}%
  \BibitemOpen
  \bibfield  {author} {\bibinfo {author} {\bibfnamefont {A.}~\bibnamefont
  {Thr\"anhardt}}, \bibinfo {author} {\bibfnamefont {C.}~\bibnamefont {Ell}},
  \bibinfo {author} {\bibfnamefont {S.}~\bibnamefont {Mosor}}, \bibinfo
  {author} {\bibfnamefont {G.}~\bibnamefont {Rupper}}, \bibinfo {author}
  {\bibfnamefont {G.}~\bibnamefont {Khitrova}}, \bibinfo {author}
  {\bibfnamefont {H.~M.}\ \bibnamefont {Gibbs}},\ and\ \bibinfo {author}
  {\bibfnamefont {S.~W.}\ \bibnamefont {Koch}},\ }\bibfield  {title} {\bibinfo
  {title} {Interplay of phonon and disorder scattering in semiconductor quantum
  wells},\ }\href {https://doi.org/10.1103/PhysRevB.68.035316} {\bibfield
  {journal} {\bibinfo  {journal} {Phys. Rev. B}\ }\textbf {\bibinfo {volume}
  {68}},\ \bibinfo {pages} {035316} (\bibinfo {year} {2003})}\BibitemShut
  {NoStop}%
\bibitem [{\citenamefont {Zhao}\ and\ \citenamefont
  {Kalt}(2004)}]{zhao2004e-ph}%
  \BibitemOpen
  \bibfield  {author} {\bibinfo {author} {\bibfnamefont {H.}~\bibnamefont
  {Zhao}}\ and\ \bibinfo {author} {\bibfnamefont {H.}~\bibnamefont {Kalt}},\
  }\bibfield  {title} {\bibinfo {title} {Direct measurement of acoustic-phonon
  scattering of hot quantum-well excitons},\ }\href
  {https://doi.org/10.1103/PhysRevB.69.233305} {\bibfield  {journal} {\bibinfo
  {journal} {Phys. Rev. B}\ }\textbf {\bibinfo {volume} {69}},\ \bibinfo
  {pages} {233305} (\bibinfo {year} {2004})}\BibitemShut {NoStop}%
\bibitem [{\citenamefont {Poltavtsev}\ \emph {et~al.}(2014)\citenamefont
  {Poltavtsev}, \citenamefont {Efimov}, \citenamefont {Dolgikh}, \citenamefont
  {Eliseev}, \citenamefont {Petrov},\ and\ \citenamefont
  {Ovsyankin}}]{poltavtsev2014e-ph}%
  \BibitemOpen
  \bibfield  {author} {\bibinfo {author} {\bibfnamefont {S.~V.}\ \bibnamefont
  {Poltavtsev}}, \bibinfo {author} {\bibfnamefont {Y.}~\bibnamefont {Efimov}},
  \bibinfo {author} {\bibfnamefont {Y.}~\bibnamefont {Dolgikh}}, \bibinfo
  {author} {\bibfnamefont {S.~A.}\ \bibnamefont {Eliseev}}, \bibinfo {author}
  {\bibfnamefont {V.~V.}\ \bibnamefont {Petrov}},\ and\ \bibinfo {author}
  {\bibfnamefont {V.~V.}\ \bibnamefont {Ovsyankin}},\ }\bibfield  {title}
  {\bibinfo {title} {Extremely low inhomogeneous broadening of exciton lines in
  shallow ({I}n,{G}a){A}s/{G}a{A}s quantum wells},\ }\href
  {https://www.sciencedirect.com/science/article/pii/S0038109814003640}
  {\bibfield  {journal} {\bibinfo  {journal} {Solid State Communications}\
  }\textbf {\bibinfo {volume} {199}},\ \bibinfo {pages} {47} (\bibinfo {year}
  {2014})}\BibitemShut {NoStop}%
\bibitem [{\citenamefont {Honold}\ \emph {et~al.}(1989)\citenamefont {Honold},
  \citenamefont {Schultheis}, \citenamefont {Kuhl},\ and\ \citenamefont
  {Tu}}]{honold1989}%
  \BibitemOpen
  \bibfield  {author} {\bibinfo {author} {\bibfnamefont {A.}~\bibnamefont
  {Honold}}, \bibinfo {author} {\bibfnamefont {L.}~\bibnamefont {Schultheis}},
  \bibinfo {author} {\bibfnamefont {J.}~\bibnamefont {Kuhl}},\ and\ \bibinfo
  {author} {\bibfnamefont {C.~W.}\ \bibnamefont {Tu}},\ }\bibfield  {title}
  {\bibinfo {title} {Collision broadening of two-dimensional excitons in a
  {G}a{A}s single quantum well},\ }\href
  {https://doi.org/10.1103/PhysRevB.40.6442} {\bibfield  {journal} {\bibinfo
  {journal} {Phys. Rev. B}\ }\textbf {\bibinfo {volume} {40}},\ \bibinfo
  {pages} {6442} (\bibinfo {year} {1989})}\BibitemShut {NoStop}%
\bibitem [{\citenamefont {Deveaud}\ \emph {et~al.}(1991)\citenamefont
  {Deveaud}, \citenamefont {Cl\'erot}, \citenamefont {Roy}, \citenamefont
  {Satzke}, \citenamefont {Sermage},\ and\ \citenamefont
  {Katzer}}]{deveaud1991}%
  \BibitemOpen
  \bibfield  {author} {\bibinfo {author} {\bibfnamefont {B.}~\bibnamefont
  {Deveaud}}, \bibinfo {author} {\bibfnamefont {F.}~\bibnamefont {Cl\'erot}},
  \bibinfo {author} {\bibfnamefont {N.}~\bibnamefont {Roy}}, \bibinfo {author}
  {\bibfnamefont {K.}~\bibnamefont {Satzke}}, \bibinfo {author} {\bibfnamefont
  {B.}~\bibnamefont {Sermage}},\ and\ \bibinfo {author} {\bibfnamefont {D.~S.}\
  \bibnamefont {Katzer}},\ }\bibfield  {title} {\bibinfo {title} {Enhanced
  radiative recombination of free excitons in {G}a{A}s quantum wells},\ }\href
  {https://doi.org/10.1103/PhysRevLett.67.2355} {\bibfield  {journal} {\bibinfo
   {journal} {Phys. Rev. Lett.}\ }\textbf {\bibinfo {volume} {67}},\ \bibinfo
  {pages} {2355} (\bibinfo {year} {1991})}\BibitemShut {NoStop}%
\bibitem [{\citenamefont {Kaindl}\ \emph {et~al.}(2003)\citenamefont {Kaindl},
  \citenamefont {Carnahan}, \citenamefont {H{\"a}gele}, \citenamefont
  {L{\"o}venich},\ and\ \citenamefont {Chemla}}]{kaindl2003}%
  \BibitemOpen
  \bibfield  {author} {\bibinfo {author} {\bibfnamefont {R.~A.}\ \bibnamefont
  {Kaindl}}, \bibinfo {author} {\bibfnamefont {M.~A.}\ \bibnamefont
  {Carnahan}}, \bibinfo {author} {\bibfnamefont {D.}~\bibnamefont
  {H{\"a}gele}}, \bibinfo {author} {\bibfnamefont {R.}~\bibnamefont
  {L{\"o}venich}},\ and\ \bibinfo {author} {\bibfnamefont {D.~S.}\ \bibnamefont
  {Chemla}},\ }\bibfield  {title} {\bibinfo {title} {Ultrafast terahertz probes
  of transient conducting and insulating phases in an electron--hole gas},\
  }\href {https://doi.org/10.1038/nature01676} {\bibfield  {journal} {\bibinfo
  {journal} {Nature}\ }\textbf {\bibinfo {volume} {423}},\ \bibinfo {pages}
  {734} (\bibinfo {year} {2003})}\BibitemShut {NoStop}%
\bibitem [{\citenamefont {Szczytko}\ \emph {et~al.}(2004)\citenamefont
  {Szczytko}, \citenamefont {Kappei}, \citenamefont {Berney}, \citenamefont
  {Morier-Genoud}, \citenamefont {Portella-Oberli},\ and\ \citenamefont
  {Deveaud}}]{szczytko2004}%
  \BibitemOpen
  \bibfield  {author} {\bibinfo {author} {\bibfnamefont {J.}~\bibnamefont
  {Szczytko}}, \bibinfo {author} {\bibfnamefont {L.}~\bibnamefont {Kappei}},
  \bibinfo {author} {\bibfnamefont {J.}~\bibnamefont {Berney}}, \bibinfo
  {author} {\bibfnamefont {F.}~\bibnamefont {Morier-Genoud}}, \bibinfo {author}
  {\bibfnamefont {M.~T.}\ \bibnamefont {Portella-Oberli}},\ and\ \bibinfo
  {author} {\bibfnamefont {B.}~\bibnamefont {Deveaud}},\ }\bibfield  {title}
  {\bibinfo {title} {Determination of the exciton formation in quantum wells
  from time-resolved interband luminescence},\ }\href
  {https://doi.org/10.1103/PhysRevLett.93.137401} {\bibfield  {journal}
  {\bibinfo  {journal} {Phys. Rev. Lett.}\ }\textbf {\bibinfo {volume} {93}},\
  \bibinfo {pages} {137401} (\bibinfo {year} {2004})}\BibitemShut {NoStop}%
\bibitem [{\citenamefont {Szczytko}\ \emph {et~al.}(2005)\citenamefont
  {Szczytko}, \citenamefont {Kappei}, \citenamefont {Berney}, \citenamefont
  {Morier-Genoud}, \citenamefont {Portella-Oberli},\ and\ \citenamefont
  {Deveaud}}]{deveaud2005PRB}%
  \BibitemOpen
  \bibfield  {author} {\bibinfo {author} {\bibfnamefont {J.}~\bibnamefont
  {Szczytko}}, \bibinfo {author} {\bibfnamefont {L.}~\bibnamefont {Kappei}},
  \bibinfo {author} {\bibfnamefont {J.}~\bibnamefont {Berney}}, \bibinfo
  {author} {\bibfnamefont {F.}~\bibnamefont {Morier-Genoud}}, \bibinfo {author}
  {\bibfnamefont {M.~T.}\ \bibnamefont {Portella-Oberli}},\ and\ \bibinfo
  {author} {\bibfnamefont {B.}~\bibnamefont {Deveaud}},\ }\bibfield  {title}
  {\bibinfo {title} {Origin of excitonic luminescence in quantum wells: Direct
  comparison of the exciton population and coulomb correlated plasma models},\
  }\href {https://doi.org/10.1103/PhysRevB.71.195313} {\bibfield  {journal}
  {\bibinfo  {journal} {Phys. Rev. B}\ }\textbf {\bibinfo {volume} {71}},\
  \bibinfo {pages} {195313} (\bibinfo {year} {2005})}\BibitemShut {NoStop}%
\bibitem [{\citenamefont {Kaindl}\ \emph {et~al.}(2009)\citenamefont {Kaindl},
  \citenamefont {H\"agele}, \citenamefont {Carnahan},\ and\ \citenamefont
  {Chemla}}]{kaindl2009}%
  \BibitemOpen
  \bibfield  {author} {\bibinfo {author} {\bibfnamefont {R.~A.}\ \bibnamefont
  {Kaindl}}, \bibinfo {author} {\bibfnamefont {D.}~\bibnamefont {H\"agele}},
  \bibinfo {author} {\bibfnamefont {M.~A.}\ \bibnamefont {Carnahan}},\ and\
  \bibinfo {author} {\bibfnamefont {D.~S.}\ \bibnamefont {Chemla}},\ }\bibfield
   {title} {\bibinfo {title} {Transient terahertz spectroscopy of excitons and
  unbound carriers in quasi-two-dimensional electron-hole gases},\ }\href
  {https://doi.org/10.1103/PhysRevB.79.045320} {\bibfield  {journal} {\bibinfo
  {journal} {Phys. Rev. B}\ }\textbf {\bibinfo {volume} {79}},\ \bibinfo
  {pages} {045320} (\bibinfo {year} {2009})}\BibitemShut {NoStop}%
\bibitem [{\citenamefont {Trifonov}\ \emph {et~al.}(2015)\citenamefont
  {Trifonov}, \citenamefont {Korotan}, \citenamefont {Kurdyubov}, \citenamefont
  {Gerlovin}, \citenamefont {Ignatiev}, \citenamefont {Efimov}, \citenamefont
  {Eliseev}, \citenamefont {Petrov}, \citenamefont {Dolgikh}, \citenamefont
  {Ovsyankin},\ and\ \citenamefont {Kavokin}}]{trifonov2015}%
  \BibitemOpen
  \bibfield  {author} {\bibinfo {author} {\bibfnamefont {A.~V.}\ \bibnamefont
  {Trifonov}}, \bibinfo {author} {\bibfnamefont {S.~N.}\ \bibnamefont
  {Korotan}}, \bibinfo {author} {\bibfnamefont {A.~S.}\ \bibnamefont
  {Kurdyubov}}, \bibinfo {author} {\bibfnamefont {I.~Y.}\ \bibnamefont
  {Gerlovin}}, \bibinfo {author} {\bibfnamefont {I.~V.}\ \bibnamefont
  {Ignatiev}}, \bibinfo {author} {\bibfnamefont {Y.~P.}\ \bibnamefont
  {Efimov}}, \bibinfo {author} {\bibfnamefont {S.~A.}\ \bibnamefont {Eliseev}},
  \bibinfo {author} {\bibfnamefont {V.~V.}\ \bibnamefont {Petrov}}, \bibinfo
  {author} {\bibfnamefont {Y.~K.}\ \bibnamefont {Dolgikh}}, \bibinfo {author}
  {\bibfnamefont {V.~V.}\ \bibnamefont {Ovsyankin}},\ and\ \bibinfo {author}
  {\bibfnamefont {A.~V.}\ \bibnamefont {Kavokin}},\ }\bibfield  {title}
  {\bibinfo {title} {Nontrivial relaxation dynamics of excitons in high-quality
  {I}n{G}a{A}s/{G}a{A}s quantum wells},\ }\href
  {https://doi.org/10.1103/PhysRevB.91.115307} {\bibfield  {journal} {\bibinfo
  {journal} {Phys. Rev. B}\ }\textbf {\bibinfo {volume} {91}},\ \bibinfo
  {pages} {115307} (\bibinfo {year} {2015})}\BibitemShut {NoStop}%
\bibitem [{\citenamefont {Beck}\ \emph {et~al.}(2016)\citenamefont {Beck},
  \citenamefont {H\"ubner}, \citenamefont {Oestreich}, \citenamefont {Bieker},
  \citenamefont {Henn}, \citenamefont {Kiessling}, \citenamefont {Ossau},\ and\
  \citenamefont {Molenkamp}}]{beck2016}%
  \BibitemOpen
  \bibfield  {author} {\bibinfo {author} {\bibfnamefont {M.}~\bibnamefont
  {Beck}}, \bibinfo {author} {\bibfnamefont {J.}~\bibnamefont {H\"ubner}},
  \bibinfo {author} {\bibfnamefont {M.}~\bibnamefont {Oestreich}}, \bibinfo
  {author} {\bibfnamefont {S.}~\bibnamefont {Bieker}}, \bibinfo {author}
  {\bibfnamefont {T.}~\bibnamefont {Henn}}, \bibinfo {author} {\bibfnamefont
  {T.}~\bibnamefont {Kiessling}}, \bibinfo {author} {\bibfnamefont
  {W.}~\bibnamefont {Ossau}},\ and\ \bibinfo {author} {\bibfnamefont {L.~W.}\
  \bibnamefont {Molenkamp}},\ }\bibfield  {title} {\bibinfo {title}
  {Thermodynamic origin of the slow free exciton photoluminescence rise in
  {G}a{A}s},\ }\href {https://doi.org/10.1103/PhysRevB.93.081204} {\bibfield
  {journal} {\bibinfo  {journal} {Phys. Rev. B}\ }\textbf {\bibinfo {volume}
  {93}},\ \bibinfo {pages} {081204} (\bibinfo {year} {2016})}\BibitemShut
  {NoStop}%
\bibitem [{\citenamefont {Kurdyubov}\ \emph {et~al.}(2021)\citenamefont
  {Kurdyubov}, \citenamefont {Trifonov}, \citenamefont {Gribakin},
  \citenamefont {Grigoryev}, \citenamefont {Gerlovin}, \citenamefont
  {Mikhailov}, \citenamefont {Ignatiev}, \citenamefont {Efimov}, \citenamefont
  {Eliseev}, \citenamefont {Lovtcius}, \citenamefont {Aßmann},\ and\
  \citenamefont {Kavokin}}]{kurdyubov2021arXiv}%
  \BibitemOpen
  \bibfield  {author} {\bibinfo {author} {\bibfnamefont {A.~S.}\ \bibnamefont
  {Kurdyubov}}, \bibinfo {author} {\bibfnamefont {A.~V.}\ \bibnamefont
  {Trifonov}}, \bibinfo {author} {\bibfnamefont {B.~F.}\ \bibnamefont
  {Gribakin}}, \bibinfo {author} {\bibfnamefont {P.~S.}\ \bibnamefont
  {Grigoryev}}, \bibinfo {author} {\bibfnamefont {I.~Y.}\ \bibnamefont
  {Gerlovin}}, \bibinfo {author} {\bibfnamefont {A.~V.}\ \bibnamefont
  {Mikhailov}}, \bibinfo {author} {\bibfnamefont {I.~V.}\ \bibnamefont
  {Ignatiev}}, \bibinfo {author} {\bibfnamefont {Y.~P.}\ \bibnamefont
  {Efimov}}, \bibinfo {author} {\bibfnamefont {S.~A.}\ \bibnamefont {Eliseev}},
  \bibinfo {author} {\bibfnamefont {V.~A.}\ \bibnamefont {Lovtcius}}, \bibinfo
  {author} {\bibfnamefont {M.}~\bibnamefont {Aßmann}},\ and\ \bibinfo {author}
  {\bibfnamefont {A.~V.}\ \bibnamefont {Kavokin}},\ }\href@noop {} {\bibinfo
  {title} {Dynamics and control of nonradiative excitons - free carriers
  mixture in {G}a{A}s/{A}l{G}a{A}s quantum wells}} (\bibinfo {year} {2021}),\
  \Eprint {https://arxiv.org/abs/2103.09867} {arXiv:2103.09867
  [cond-mat.mes-hall]} \BibitemShut {NoStop}%
\bibitem [{\citenamefont {Berger}\ \emph {et~al.}(2020)\citenamefont {Berger},
  \citenamefont {Schmidt}, \citenamefont {Ma}, \citenamefont {Schumacher},
  \citenamefont {Schneider}, \citenamefont {H\"ofling},\ and\ \citenamefont
  {A\ss{}mann}}]{assmann2020}%
  \BibitemOpen
  \bibfield  {author} {\bibinfo {author} {\bibfnamefont {B.}~\bibnamefont
  {Berger}}, \bibinfo {author} {\bibfnamefont {D.}~\bibnamefont {Schmidt}},
  \bibinfo {author} {\bibfnamefont {X.}~\bibnamefont {Ma}}, \bibinfo {author}
  {\bibfnamefont {S.}~\bibnamefont {Schumacher}}, \bibinfo {author}
  {\bibfnamefont {C.}~\bibnamefont {Schneider}}, \bibinfo {author}
  {\bibfnamefont {S.}~\bibnamefont {H\"ofling}},\ and\ \bibinfo {author}
  {\bibfnamefont {M.}~\bibnamefont {A\ss{}mann}},\ }\bibfield  {title}
  {\bibinfo {title} {Formation dynamics of exciton-polariton vortices created
  by nonresonant annular pumping},\ }\href
  {https://doi.org/10.1103/PhysRevB.101.245309} {\bibfield  {journal} {\bibinfo
   {journal} {Phys. Rev. B}\ }\textbf {\bibinfo {volume} {101}},\ \bibinfo
  {pages} {245309} (\bibinfo {year} {2020})}\BibitemShut {NoStop}%
\bibitem [{\citenamefont {Deveaud}\ \emph {et~al.}(2005)\citenamefont
  {Deveaud}, \citenamefont {Kappei}, \citenamefont {Berney}, \citenamefont
  {Morier-Genoud}, \citenamefont {Portella-Oberli}, \citenamefont {Szczytko},\
  and\ \citenamefont {Piermarocchi}}]{deveaud2005chemPhys}%
  \BibitemOpen
  \bibfield  {author} {\bibinfo {author} {\bibfnamefont {B.}~\bibnamefont
  {Deveaud}}, \bibinfo {author} {\bibfnamefont {L.}~\bibnamefont {Kappei}},
  \bibinfo {author} {\bibfnamefont {J.}~\bibnamefont {Berney}}, \bibinfo
  {author} {\bibfnamefont {F.}~\bibnamefont {Morier-Genoud}}, \bibinfo {author}
  {\bibfnamefont {M.}~\bibnamefont {Portella-Oberli}}, \bibinfo {author}
  {\bibfnamefont {J.}~\bibnamefont {Szczytko}},\ and\ \bibinfo {author}
  {\bibfnamefont {C.}~\bibnamefont {Piermarocchi}},\ }\bibfield  {title}
  {\bibinfo {title} {Excitonic effects in the luminescence of quantum wells},\
  }\href {https://doi.org/https://doi.org/10.1016/j.chemphys.2005.06.045}
  {\bibfield  {journal} {\bibinfo  {journal} {Chemical Physics}\ }\textbf
  {\bibinfo {volume} {318}},\ \bibinfo {pages} {104} (\bibinfo {year}
  {2005})}\BibitemShut {NoStop}%
\bibitem [{\citenamefont {Haug}(1976)}]{haug1976}%
  \BibitemOpen
  \bibfield  {author} {\bibinfo {author} {\bibfnamefont {H.}~\bibnamefont
  {Haug}},\ }\bibfield  {title} {\bibinfo {title} {On the phase transitions for
  the electronic excitations in semiconductors},\ }\href
  {https://doi.org/10.1007/BF01351524} {\bibfield  {journal} {\bibinfo
  {journal} {Zeitschrift f{\"u}r Physik B Condensed Matter}\ }\textbf {\bibinfo
  {volume} {24}},\ \bibinfo {pages} {351} (\bibinfo {year} {1976})}\BibitemShut
  {NoStop}%
\bibitem [{\citenamefont {Feng}\ and\ \citenamefont
  {Spector}(1987)}]{feng1987}%
  \BibitemOpen
  \bibfield  {author} {\bibinfo {author} {\bibfnamefont {Y.-P.}\ \bibnamefont
  {Feng}}\ and\ \bibinfo {author} {\bibfnamefont {H.~N.}\ \bibnamefont
  {Spector}},\ }\bibfield  {title} {\bibinfo {title} {Scattering of excitons by
  free carriers in semiconducting quantum well structures},\ }\href
  {https://doi.org/https://doi.org/10.1016/0022-3697(87)90146-6} {\bibfield
  {journal} {\bibinfo  {journal} {Journal of Physics and Chemistry of Solids}\
  }\textbf {\bibinfo {volume} {48}},\ \bibinfo {pages} {593} (\bibinfo {year}
  {1987})}\BibitemShut {NoStop}%
\bibitem [{\citenamefont {Hiroshima}(1989)}]{hiroshima1989}%
  \BibitemOpen
  \bibfield  {author} {\bibinfo {author} {\bibfnamefont {T.}~\bibnamefont
  {Hiroshima}},\ }\bibfield  {title} {\bibinfo {title} {Nonresonant excitonic
  optical nonlinearity in semiconductors},\ }\href
  {https://doi.org/10.1103/PhysRevB.40.3862} {\bibfield  {journal} {\bibinfo
  {journal} {Phys. Rev. B}\ }\textbf {\bibinfo {volume} {40}},\ \bibinfo
  {pages} {3862} (\bibinfo {year} {1989})}\BibitemShut {NoStop}%
\bibitem [{\citenamefont {May}\ \emph {et~al.}(1985)\citenamefont {May},
  \citenamefont {Boldt},\ and\ \citenamefont {Henneberger}}]{dense_gas_1}%
  \BibitemOpen
  \bibfield  {author} {\bibinfo {author} {\bibfnamefont {V.}~\bibnamefont
  {May}}, \bibinfo {author} {\bibfnamefont {F.}~\bibnamefont {Boldt}},\ and\
  \bibinfo {author} {\bibfnamefont {K.}~\bibnamefont {Henneberger}},\
  }\bibfield  {title} {\bibinfo {title} {Many-body theory for the dense exciton
  gas of direct semiconductors {I}. {G}eneral considerations},\ }\href
  {https://doi.org/https://doi.org/10.1002/pssb.2221290232} {\bibfield
  {journal} {\bibinfo  {journal} {physica status solidi (b)}\ }\textbf
  {\bibinfo {volume} {129}},\ \bibinfo {pages} {717} (\bibinfo {year}
  {1985})}\BibitemShut {NoStop}%
\bibitem [{\citenamefont {Boldt}\ \emph {et~al.}(1985)\citenamefont {Boldt},
  \citenamefont {Henneberger},\ and\ \citenamefont {May}}]{dense_gas_2}%
  \BibitemOpen
  \bibfield  {author} {\bibinfo {author} {\bibfnamefont {F.}~\bibnamefont
  {Boldt}}, \bibinfo {author} {\bibfnamefont {K.}~\bibnamefont {Henneberger}},\
  and\ \bibinfo {author} {\bibfnamefont {V.}~\bibnamefont {May}},\ }\bibfield
  {title} {\bibinfo {title} {Many-body theory for the dense exciton gas of
  direct semiconductors {II}. {C}alculation of exciton level shift and damping
  in dependence on exciton density},\ }\href
  {https://doi.org/https://doi.org/10.1002/pssb.2221300231} {\bibfield
  {journal} {\bibinfo  {journal} {physica status solidi (b)}\ }\textbf
  {\bibinfo {volume} {130}},\ \bibinfo {pages} {675} (\bibinfo {year}
  {1985})}\BibitemShut {NoStop}%
\bibitem [{\citenamefont {Amand}\ \emph {et~al.}(1997)\citenamefont {Amand},
  \citenamefont {Robart}, \citenamefont {Marie}, \citenamefont {Brousseau},
  \citenamefont {Le~Jeune},\ and\ \citenamefont {Barrau}}]{amand1997}%
  \BibitemOpen
  \bibfield  {author} {\bibinfo {author} {\bibfnamefont {T.}~\bibnamefont
  {Amand}}, \bibinfo {author} {\bibfnamefont {D.}~\bibnamefont {Robart}},
  \bibinfo {author} {\bibfnamefont {X.}~\bibnamefont {Marie}}, \bibinfo
  {author} {\bibfnamefont {M.}~\bibnamefont {Brousseau}}, \bibinfo {author}
  {\bibfnamefont {P.}~\bibnamefont {Le~Jeune}},\ and\ \bibinfo {author}
  {\bibfnamefont {J.}~\bibnamefont {Barrau}},\ }\bibfield  {title} {\bibinfo
  {title} {Spin relaxation in polarized interacting exciton gas in quantum
  wells},\ }\href {https://doi.org/10.1103/PhysRevB.55.9880} {\bibfield
  {journal} {\bibinfo  {journal} {Phys. Rev. B}\ }\textbf {\bibinfo {volume}
  {55}},\ \bibinfo {pages} {9880} (\bibinfo {year} {1997})}\BibitemShut
  {NoStop}%
\bibitem [{\citenamefont {Ciuti}\ \emph {et~al.}(1998)\citenamefont {Ciuti},
  \citenamefont {Savona}, \citenamefont {Piermarocchi}, \citenamefont
  {Quattropani},\ and\ \citenamefont {Schwendimann}}]{ciuti1998}%
  \BibitemOpen
  \bibfield  {author} {\bibinfo {author} {\bibfnamefont {C.}~\bibnamefont
  {Ciuti}}, \bibinfo {author} {\bibfnamefont {V.}~\bibnamefont {Savona}},
  \bibinfo {author} {\bibfnamefont {C.}~\bibnamefont {Piermarocchi}}, \bibinfo
  {author} {\bibfnamefont {A.}~\bibnamefont {Quattropani}},\ and\ \bibinfo
  {author} {\bibfnamefont {P.}~\bibnamefont {Schwendimann}},\ }\bibfield
  {title} {\bibinfo {title} {Role of the exchange of carriers in elastic
  exciton-exciton scattering in quantum wells},\ }\href
  {https://doi.org/10.1103/PhysRevB.58.7926} {\bibfield  {journal} {\bibinfo
  {journal} {Phys. Rev. B}\ }\textbf {\bibinfo {volume} {58}},\ \bibinfo
  {pages} {7926} (\bibinfo {year} {1998})}\BibitemShut {NoStop}%
\bibitem [{\citenamefont {Ramon}\ \emph {et~al.}(2003)\citenamefont {Ramon},
  \citenamefont {Mann},\ and\ \citenamefont
  {Cohen}}]{cohen_neutral&charged_exciton-electron2003}%
  \BibitemOpen
  \bibfield  {author} {\bibinfo {author} {\bibfnamefont {G.}~\bibnamefont
  {Ramon}}, \bibinfo {author} {\bibfnamefont {A.}~\bibnamefont {Mann}},\ and\
  \bibinfo {author} {\bibfnamefont {E.}~\bibnamefont {Cohen}},\ }\bibfield
  {title} {\bibinfo {title} {Theory of neutral and charged exciton scattering
  with electrons in semiconductor quantum wells},\ }\href
  {https://doi.org/10.1103/PhysRevB.67.045323} {\bibfield  {journal} {\bibinfo
  {journal} {Phys. Rev. B}\ }\textbf {\bibinfo {volume} {67}},\ \bibinfo
  {pages} {045323} (\bibinfo {year} {2003})}\BibitemShut {NoStop}%
\bibitem [{\citenamefont {Ouerdane}\ \emph {et~al.}(2008)\citenamefont
  {Ouerdane}, \citenamefont {Varache}, \citenamefont {Portnoi},\ and\
  \citenamefont {Galbraith}}]{ouerdane2008}%
  \BibitemOpen
  \bibfield  {author} {\bibinfo {author} {\bibfnamefont {H.}~\bibnamefont
  {Ouerdane}}, \bibinfo {author} {\bibfnamefont {R.}~\bibnamefont {Varache}},
  \bibinfo {author} {\bibfnamefont {M.~E.}\ \bibnamefont {Portnoi}},\ and\
  \bibinfo {author} {\bibfnamefont {I.}~\bibnamefont {Galbraith}},\ }\bibfield
  {title} {\bibinfo {title} {Photon emission induced by elastic exciton-carrier
  scattering in semiconductor quantum wells},\ }\href
  {https://doi.org/10.1140/epjb/e2008-00355-x} {\bibfield  {journal} {\bibinfo
  {journal} {The European Physical Journal B}\ }\textbf {\bibinfo {volume}
  {65}},\ \bibinfo {pages} {195} (\bibinfo {year} {2008})}\BibitemShut
  {NoStop}%
\bibitem [{\citenamefont {Schindler}\ and\ \citenamefont
  {Zimmermann}(2008)}]{schindler2008}%
  \BibitemOpen
  \bibfield  {author} {\bibinfo {author} {\bibfnamefont {C.}~\bibnamefont
  {Schindler}}\ and\ \bibinfo {author} {\bibfnamefont {R.}~\bibnamefont
  {Zimmermann}},\ }\bibfield  {title} {\bibinfo {title} {Analysis of the
  exciton-exciton interaction in semiconductor quantum wells},\ }\href
  {https://doi.org/10.1103/PhysRevB.78.045313} {\bibfield  {journal} {\bibinfo
  {journal} {Phys. Rev. B}\ }\textbf {\bibinfo {volume} {78}},\ \bibinfo
  {pages} {045313} (\bibinfo {year} {2008})}\BibitemShut {NoStop}%
\bibitem [{\citenamefont {Okumura}\ and\ \citenamefont
  {Ogawa}(2001)}]{okumura_ogawa2001bosonization}%
  \BibitemOpen
  \bibfield  {author} {\bibinfo {author} {\bibfnamefont {S.}~\bibnamefont
  {Okumura}}\ and\ \bibinfo {author} {\bibfnamefont {T.}~\bibnamefont
  {Ogawa}},\ }\bibfield  {title} {\bibinfo {title} {Boson representation of
  two-exciton correlations: An exact treatment of composite-particle effects},\
  }\href {https://doi.org/10.1103/PhysRevB.65.035105} {\bibfield  {journal}
  {\bibinfo  {journal} {Phys. Rev. B}\ }\textbf {\bibinfo {volume} {65}},\
  \bibinfo {pages} {035105} (\bibinfo {year} {2001})}\BibitemShut {NoStop}%
\bibitem [{\citenamefont {Combescot}\ \emph {et~al.}(2008)\citenamefont
  {Combescot}, \citenamefont {Betbeder-Matibet},\ and\ \citenamefont
  {Dubin}}]{combescot2008}%
  \BibitemOpen
  \bibfield  {author} {\bibinfo {author} {\bibfnamefont {M.}~\bibnamefont
  {Combescot}}, \bibinfo {author} {\bibfnamefont {O.}~\bibnamefont
  {Betbeder-Matibet}},\ and\ \bibinfo {author} {\bibfnamefont {F.}~\bibnamefont
  {Dubin}},\ }\bibfield  {title} {\bibinfo {title} {The many-body physics of
  composite bosons},\ }\href
  {https://www.sciencedirect.com/science/article/pii/S0370157308000975}
  {\bibfield  {journal} {\bibinfo  {journal} {Physics Reports}\ }\textbf
  {\bibinfo {volume} {463}},\ \bibinfo {pages} {215} (\bibinfo {year}
  {2008})}\BibitemShut {NoStop}%
\bibitem [{\citenamefont {Glazov}\ \emph {et~al.}(2009)\citenamefont {Glazov},
  \citenamefont {Ouerdane}, \citenamefont {Pilozzi}, \citenamefont {Malpuech},
  \citenamefont {Kavokin},\ and\ \citenamefont {D'Andrea}}]{glazov2009}%
  \BibitemOpen
  \bibfield  {author} {\bibinfo {author} {\bibfnamefont {M.~M.}\ \bibnamefont
  {Glazov}}, \bibinfo {author} {\bibfnamefont {H.}~\bibnamefont {Ouerdane}},
  \bibinfo {author} {\bibfnamefont {L.}~\bibnamefont {Pilozzi}}, \bibinfo
  {author} {\bibfnamefont {G.}~\bibnamefont {Malpuech}}, \bibinfo {author}
  {\bibfnamefont {A.~V.}\ \bibnamefont {Kavokin}},\ and\ \bibinfo {author}
  {\bibfnamefont {A.}~\bibnamefont {D'Andrea}},\ }\bibfield  {title} {\bibinfo
  {title} {Polariton-polariton scattering in microcavities: A microscopic
  theory},\ }\href {https://doi.org/10.1103/PhysRevB.80.155306} {\bibfield
  {journal} {\bibinfo  {journal} {Phys. Rev. B}\ }\textbf {\bibinfo {volume}
  {80}},\ \bibinfo {pages} {155306} (\bibinfo {year} {2009})}\BibitemShut
  {NoStop}%
\bibitem [{\citenamefont {Grigoryev}(2021)}]{grigoryev2021unpublished}%
  \BibitemOpen
  \bibfield  {author} {\bibinfo {author} {\bibfnamefont {P.~S.}\ \bibnamefont
  {Grigoryev}}} (\bibinfo {year} {2021}),\ \bibinfo {note}
  {unpublished}\BibitemShut {NoStop}%
\bibitem [{\citenamefont {Kressner}(2005)}]{kressner2005KrylovSchur}%
  \BibitemOpen
  \bibfield  {author} {\bibinfo {author} {\bibfnamefont {D.}~\bibnamefont
  {Kressner}},\ }\bibinfo {title} {The krylov-schur algorithm},\ in\ \href
  {https://doi.org/10.1007/3-540-28502-4_3} {\emph {\bibinfo {booktitle}
  {Numerical Methods for General and Structured Eigenvalue Problems}}}\
  (\bibinfo  {publisher} {Springer Berlin Heidelberg},\ \bibinfo {address}
  {Berlin, Heidelberg},\ \bibinfo {year} {2005})\ pp.\ \bibinfo {pages}
  {113--130}\BibitemShut {NoStop}%
\bibitem [{\citenamefont {Davies}(1997)}]{davies1997book}%
  \BibitemOpen
  \bibfield  {author} {\bibinfo {author} {\bibfnamefont {J.~H.}\ \bibnamefont
  {Davies}},\ }\href {https://doi.org/10.1017/CBO9780511819070} {\emph
  {\bibinfo {title} {The Physics of Low-dimensional Semiconductors: An
  Introduction}}}\ (\bibinfo  {publisher} {Cambridge University Press},\
  \bibinfo {year} {1997})\BibitemShut {NoStop}%
\bibitem [{\citenamefont {Blackwood}\ \emph {et~al.}(1994)\citenamefont
  {Blackwood}, \citenamefont {Snelling}, \citenamefont {Harley}, \citenamefont
  {Andrews},\ and\ \citenamefont {Foxon}}]{blackwood1994_e-h_exchange}%
  \BibitemOpen
  \bibfield  {author} {\bibinfo {author} {\bibfnamefont {E.}~\bibnamefont
  {Blackwood}}, \bibinfo {author} {\bibfnamefont {M.~J.}\ \bibnamefont
  {Snelling}}, \bibinfo {author} {\bibfnamefont {R.~T.}\ \bibnamefont
  {Harley}}, \bibinfo {author} {\bibfnamefont {S.~R.}\ \bibnamefont
  {Andrews}},\ and\ \bibinfo {author} {\bibfnamefont {C.~T.~B.}\ \bibnamefont
  {Foxon}},\ }\bibfield  {title} {\bibinfo {title} {Exchange interaction of
  excitons in gaas heterostructures},\ }\href
  {https://doi.org/10.1103/PhysRevB.50.14246} {\bibfield  {journal} {\bibinfo
  {journal} {Phys. Rev. B}\ }\textbf {\bibinfo {volume} {50}},\ \bibinfo
  {pages} {14246} (\bibinfo {year} {1994})}\BibitemShut {NoStop}%
\bibitem [{\citenamefont {Landau}\ and\ \citenamefont
  {Lifshitz}(2013)}]{landafshitz}%
  \BibitemOpen
  \bibfield  {author} {\bibinfo {author} {\bibfnamefont {L.~D.}\ \bibnamefont
  {Landau}}\ and\ \bibinfo {author} {\bibfnamefont {E.~M.}\ \bibnamefont
  {Lifshitz}},\ }\href@noop {} {\emph {\bibinfo {title} {Course of theoretical
  physics}}}\ (\bibinfo  {publisher} {Elsevier},\ \bibinfo {year}
  {2013})\BibitemShut {NoStop}%
\bibitem [{\citenamefont {Trifonov}\ \emph {et~al.}(2019)\citenamefont
  {Trifonov}, \citenamefont {Khramtsov}, \citenamefont {Kavokin}, \citenamefont
  {Ignatiev}, \citenamefont {Kavokin}, \citenamefont {Efimov}, \citenamefont
  {Eliseev}, \citenamefont {Shapochkin},\ and\ \citenamefont
  {Bayer}}]{trifonov2019PRL}%
  \BibitemOpen
  \bibfield  {author} {\bibinfo {author} {\bibfnamefont {A.~V.}\ \bibnamefont
  {Trifonov}}, \bibinfo {author} {\bibfnamefont {E.~S.}\ \bibnamefont
  {Khramtsov}}, \bibinfo {author} {\bibfnamefont {K.~V.}\ \bibnamefont
  {Kavokin}}, \bibinfo {author} {\bibfnamefont {I.~V.}\ \bibnamefont
  {Ignatiev}}, \bibinfo {author} {\bibfnamefont {A.~V.}\ \bibnamefont
  {Kavokin}}, \bibinfo {author} {\bibfnamefont {Y.~P.}\ \bibnamefont {Efimov}},
  \bibinfo {author} {\bibfnamefont {S.~A.}\ \bibnamefont {Eliseev}}, \bibinfo
  {author} {\bibfnamefont {P.~Y.}\ \bibnamefont {Shapochkin}},\ and\ \bibinfo
  {author} {\bibfnamefont {M.}~\bibnamefont {Bayer}},\ }\bibfield  {title}
  {\bibinfo {title} {Nanosecond spin coherence time of nonradiative excitons in
  {G}a{A}s/{A}l{G}a{A}s quantum wells},\ }\href
  {https://doi.org/10.1103/PhysRevLett.122.147401} {\bibfield  {journal}
  {\bibinfo  {journal} {Phys. Rev. Lett.}\ }\textbf {\bibinfo {volume} {122}},\
  \bibinfo {pages} {147401} (\bibinfo {year} {2019})}\BibitemShut {NoStop}%
\bibitem [{\citenamefont {Vurgaftman}\ \emph {et~al.}(2001)\citenamefont
  {Vurgaftman}, \citenamefont {Meyer},\ and\ \citenamefont
  {Ram-Mohan}}]{Vurgaftman2001}%
  \BibitemOpen
  \bibfield  {author} {\bibinfo {author} {\bibfnamefont {I.}~\bibnamefont
  {Vurgaftman}}, \bibinfo {author} {\bibfnamefont {J.~R.}\ \bibnamefont
  {Meyer}},\ and\ \bibinfo {author} {\bibfnamefont {L.~R.}\ \bibnamefont
  {Ram-Mohan}},\ }\bibfield  {title} {\bibinfo {title} {Band parameters for
  iii–v compound semiconductors and their alloys},\ }\href
  {https://doi.org/10.1063/1.1368156} {\bibfield  {journal} {\bibinfo
  {journal} {Journal of Applied Physics}\ }\textbf {\bibinfo {volume} {89}},\
  \bibinfo {pages} {5815} (\bibinfo {year} {2001})}\BibitemShut {NoStop}%
\bibitem [{sou()}]{source_code}%
  \BibitemOpen
  \href@noop {} {}\bibinfo {note} {The source code can be provided upon
  reasonable request.}\BibitemShut {Stop}%
\bibitem [{Note1()}]{Note1}%
  \BibitemOpen
  \bibinfo {note} {This may seem contradictory to us neglecting scattering to
  excited states, however, the obtained broadening is in the order of 1~meV.
  This is not enough for the possible transitions described in Sec. \ref
  {subsec:scatteringAmplitudes}}\BibitemShut {NoStop}%
\bibitem [{\citenamefont {Gasiorowicz}(2007)}]{gasiorowicz2007quantum}%
  \BibitemOpen
  \bibfield  {author} {\bibinfo {author} {\bibfnamefont {S.}~\bibnamefont
  {Gasiorowicz}},\ }\href@noop {} {\emph {\bibinfo {title} {Quantum physics}}}\
  (\bibinfo  {publisher} {John Wiley \& Sons},\ \bibinfo {year}
  {2007})\BibitemShut {NoStop}%
\bibitem [{\citenamefont {Balkan}\ \emph {et~al.}(1989)\citenamefont {Balkan},
  \citenamefont {Ridley}, \citenamefont {Emeny},\ and\ \citenamefont
  {Goodridge}}]{balkan1989}%
  \BibitemOpen
  \bibfield  {author} {\bibinfo {author} {\bibfnamefont {N.}~\bibnamefont
  {Balkan}}, \bibinfo {author} {\bibfnamefont {B.~K.}\ \bibnamefont {Ridley}},
  \bibinfo {author} {\bibfnamefont {M.}~\bibnamefont {Emeny}},\ and\ \bibinfo
  {author} {\bibfnamefont {I.}~\bibnamefont {Goodridge}},\ }\bibfield  {title}
  {\bibinfo {title} {Hot-electron energy relaxation rates in {GaAs}/{GaAlAs}
  quantum wells},\ }\href {https://doi.org/10.1088/0268-1242/4/10/004}
  {\bibfield  {journal} {\bibinfo  {journal} {Semiconductor Science and
  Technology}\ }\textbf {\bibinfo {volume} {4}},\ \bibinfo {pages} {852}
  (\bibinfo {year} {1989})}\BibitemShut {NoStop}%
\end{thebibliography}

%

\end{document}